\documentclass{aa}  

\usepackage{caption}
\usepackage{subcaption}
\usepackage{epstopdf}
\usepackage{amsmath}
\usepackage[english]{babel}
\usepackage{amssymb}
\usepackage{wrapfig}
\usepackage{pdflscape}
\usepackage{lscape}
\usepackage{longtable}
\usepackage{appendix}
\usepackage{xcolor}
\usepackage{textcomp}
\usepackage{multirow}
\bibpunct{(}{)}{;}{a}{}{,}
\usepackage{array}
\usepackage{ragged2e}
\usepackage{adjustbox}
\usepackage{placeins}
\usepackage{soul}
\usepackage{float}

\usepackage{makecell}
\usepackage{graphicx}
\usepackage{natbib}
\usepackage{txfonts}
\usepackage{hyperref}
\begin{document} 

   \title{An X-ray and optical spectral study of the changing-look narrow-line Seyfert 1 2MASX J0413-0050}

   \author{A. Vietri
        \inst{1,2,3}\thanks{amelia.vietri@phd.unipd.it}
          \and
          A. Tortosa\inst{4}\thanks{alessia.tortosa@inaf.it}
          \and
          D. Ili\'c \inst{5,6}
          \and
          S. Ciroi \inst{1}
          \and 
          M. Berton\inst{3}
          \and
          E. J\"arvel\"a \inst{7}
          \and
          C. Ricci \inst{8,9,10} 
          \and 
          E. Sani \inst{3}
          \and
          L. Crepaldi  \inst{1,11}
          \and
          B. Dalla Barba \inst{11,12}
          \and
          S. Chen  \inst{13}
          \and 
          E. Congiu \inst{3}
          \and
          P. Condò \inst{14}
          \and
           I. Varglund{} \inst{15,16,17}
          \and
          G. Rodighiero \inst{1,2}          
            }
          
   \institute{
   $^1$ Dipartimento di Fisica e Astronomia ``G. Galilei", Universit\`a di Padova, Vicolo dell'Osservatorio 3, 35122, Padova, Italy; \\
   $^2$ INAF – Astronomical Observatory of Padova, Vicolo dell'Osservatorio 5, 35122, Padova, Italy; \\
   $^3$ European Southern Observatory, Alonso de C\'ordova 3107, Casilla 19, Santiago 19001, Chile;\\
   $^4$ INAF – Astrophysics and Space Science Observatory of Bologna, Via Gobetti, 93/3,  40129 Bologna, Italy;\\
   $^5$ University of Belgrade - Faculty of Mathematics, Department of astronomy, Studentski trg 16, 11000 Belgrade, Serbia;\\
   $^6$ Hamburger Sternwarte, Universit\"at Hamburg, Gojenbergsweg 112, 21029 Hamburg, Germany;\\
   $^7$ Department of Physics \& Astronomy, Texas Tech University, Box 41051, Lubbock, TX, 79409-1051, USA; \\
   $^8$ Department of Astronomy, University of Geneva, ch. d'Ecogia 16, 1290, Versoix, Switzerland; \\
   $^9$ Instituto de Estudios Astrof\'isicos, Facultad de Ingenier\'ia y Ciencias, Universidad Diego Portales, Avenida Ejercito Libertador 441,Santiago, Chile;\\
   $^{10}$ Kavli Institute for Astronomy and Astrophysics, Peking University, Beijing 100871, China;\\
   $^{11}$ Osservatorio Astronomico di Brera, Istituto Nazionale di Astrofisica (INAF), Via Emilio Bianchi, 46, 23807 Merate, Italy;\\
   $^{12}$ Dipartimento di Scienza e Alta Tecnologia, Università dell'Insubria, Via Valleggio 11, 22100 Como, Italy;\\
   $^{13}$ Physics Department, Technion, Haifa 32000, Israel;\\
   $^{14}$ Dipartimento di Fisica, Università di Roma Tor Vergata, Via della Ricerca Scientifica, 1, Roma 00133, Italy;\\
   $^{15}$ Centre for Astrophysics Research, University of Hertfordshire, College Lane, Hatfield AL10 9AB, UK; \\
   $^{16}$ Aalto University Metsähovi Radio Observatory, Metsähovintie 114, FI-02540 Kylmälä, Finland;\\
   $^{17}$ Aalto University Department of Electronics and Nanoengineering, P.O. Box 15500, FI-00076 AALTO, Finland.}

   \date{Received ...; accepted ...}

% \abstract{}{}{}{}{} 
% 5 {} token are mandatory
 
\newcommand{\kms}{km~s$^{-1}$}
\newcommand{\fluxcgs}{ergs~s$^{-1}$~cm$^{-2}$}
\newcommand{\lumcgs}{ergs~s$^{-1}$}
\newcommand{\xmm}{{\it XMM}-Newton}
\newcommand{\swift}{{\it Swift}/XRT}
\newcommand{\nustar}{\textit{NuSTAR}}
\newcommand{\nicer}{\textit{NICER}}
\newcommand*\red[1]{\textcolor{red}{#1}}
\newcommand{\ergs}{erg s$^{-1}$}
\newcommand{\chired}{$\chi^2_\nu$}

  \abstract
   {Active galactic nuclei (AGN) showing dramatic spectral and flux variations, either due to changes in the accretion rate (\textit{changing-state} CS-AGN) of the supermassive black hole or in the line-of-sight column density (\textit{changing-obscuration} CO-AGN), have been classified as changing-look (CL) AGN. Here we present a peculiar source, 2MASX J0413-0050, first identified as a narrow-line Seyfert 1 (NLS1s) galaxy in 2004. When re-observed twice in 2021, it showed a transition in the spectral type (towards a Seyfert 1.9) and the complete and mysterious disappearance of the H$\beta$ line while source was in a high‑accretion state. In the meantime, the X-ray flux decreased between observations taken in 2020 and 2022, and again in the most recent spectrum of 2023. Shortly after this, another optical spectrum revealed the re‑emergence of both the narrow and broad H$\beta$ components (Seyfert 1.8). Despite the fact that it was not possible to retrieve the line-of-sight column density from the X-ray spectra, which would have helped in assessing whether this event could be attributed to a CO‑AGN scenario, the observational evidence does not necessarily support such an interpretation. J0413-0050 may have undergone several switch-on/switch-off phases over the past 20 years, on an unknown timescale, which could have affected the accretion power and, consequently, the optical continuum and so the emission lines coming from the broad-line region (BLR). For these reasons, it is reasonable to classify this source as a CS-AGN. The case of J0413-0050 supports the hypothesis that NLS1s can indeed experience CL phenomena.}
        
   \keywords{Galaxies: active - Galaxies: Seyfert -Xrays: individuals: 2MASX J04130709-0050165}

   \maketitle

%-------------------------------------------------------------------
\section{Introduction}
\label{sec:intro}
Observational evidence has proved that active galactic nuclei (AGN) are powered by accreting supermassive black holes (SMBHs, $M_{\rm BH}$ $\approx$ 10$^6$-$10^{10}$ M$_\odot$) whose growth is regulated by an active feeding process from their surroundings \citep[e.g.][]{Magorrian98}. Several diverse classes of AGN are known, whose classification is often based on multi-band spectral features, showing different observable properties mainly due to the orientation with respect to the observer \citep{Antonucci93}. The physical reason behind the main observational difference relies on the presence of an obscuring dusty medium, probably lying on the plane of the accretion disc, that it was originally hypothesised to have a toroidal shape. However, an increasing number of studies favoured a more complex clumpy structure for this absorbing medium \citep{Elitzur06}, for which a large range of dust temperatures can be found coexisting at similar distances from the source \citep{Nenkova08b}.

In type 2 AGN, this medium prevents the ionised light coming from the broad-line region (BLR) from reaching the observer and, consequently, hides the presence of the broad lines ($\gtrsim$ 1000 \kms) in the optical spectra. Conversely, when the AGN is observed from a small viewing angle (face-on view), the line-of-sight does not intercept the dusty medium and the AGN, classified as Type~1 (unobscured), exhibits both broad and narrow ($\lesssim$ 1000 \kms) optical emission lines in its spectra. A finer classification also exists \citep{Osterbrock77} for sources sharing properties of Type~1 and Type~2, the Intermediate Seyfert (IS) \citep{Dallabarba23}, classified based on the increasing faintness of the broad component with respect to the narrow one \citep[Type~1.2,~1.5,~1.8 and~1.9,][]{Vaona12}. This model of classification can also be matched with the X-ray observational properties, identifying Type~1 AGN with unobscured sources showing line-of-sight column density $N_{\rm H} \leq 10^{22} \rm cm^{-2}$ and, Type~2 AGN with obscured sources having $N_{\rm H} \geq 10^{22} \rm cm^{-2}$ \citep[e.g.,][]{Ricci17}. 

AGN variability, spanning over all frequencies and timescales, is one of their defining elements. When variability becomes drastic to the point that the AGN can switch from one classification type to another over short time scales \citep[less than a few months; e.g.][]{Zeltyn24}, the observed spectral changes cannot be explained in the context of the unification model. Starting from the first evidence in the 1970s \citep{Pastoriza70, Tohline76, Barr77}, an increasing number of sources have been observed showing this kind of drastic changes in the optical/UV and X-rays spectral properties, requiring the definition of a complete new class: changing-look (CL) AGN.

This new classification is primarily associated with the extreme changes in the X-ray absorption of the AGN, bringing the line-of-sight column density from Compton thin ($N_{\rm H}$ $\leq 10^{-24} \rm cm^{-2}$) to Compton thick (CT, i.e. $N_{\rm H}$ $> 10^{-24} \rm cm^{-2}$, \citealp{Matt03, Ricci15}) and vice versa, due to a strong variation in the obscuring material covering the X-ray continuum source (e.g. \citealp{Matt03, Ricci16}). 
Recently, this change in the X-ray spectrum has been labelled as \textit{changing-obscuration} (CO-AGN) phenomenon. These temporal variations in $N_{\rm H}$ are generally observed in the X-rays since the size of the region producing this high energy emission, the corona, is significantly smaller than the BLR. However, the obscuring clumpy material can move, causing variability \citep{Yaqoob89}, and it can go through the BLR or even beyond the dust sublimation radius, reaching the torus \citep{Ricci23}. As the level of the obscuration increases and the source becomes CT, the X-ray continuum is suppressed by the effect of the Compton scattering and the features of the reprocessed X-rays (Fe K$\alpha$ line, Compton hump) become prominent in the spectrum. 

Conversely, drastic variability in the optical/UV range can be due to a change in the accretion flow, which in turn causes variability in the continuum and the appearance or disappearance of the broad emission lines \citep{Graham20}. This is known as a \textit{changing-state} (CS-AGN), and can be as dramatic as a spectral change from Type~1 to Type~2, and vice versa \citep{Lamassa15}, falling under the name of "turn-off" and "turn-on" events \citep{Arevalo24, Sanchez24}, respectively. Several explanations have been proposed involving processes occurring in the accretion disc, such as thermal instabilities \citep{Grupe15, Stern18, Sniegowska22}, tidal disruptions of stars into AGN discs \citep[e.g.,][]{Merloni15}, rapid mass-accretion rate \citep{Noda18} drops and other disc events. CS transitions are therefore an important probe of the accretion disc physics, allowing us to study the different regions emitting continuum and generating the emission lines. CS-AGN need to lie close to the accretion state transition threshold, beyond which the ionising luminosity changes and the emission lines are produced, to account for the variable accretion rate scenario \citep{Li22, Zeltyn24, Wang24}. Indeed, spectral-types changes usually happen in AGN which show bolometric luminosities of a few percent of Eddington luminosity, $L_{\rm Edd}$ \citep{Noda18}.
 
Although AGN with a high bolometric luminosity $L_{\rm bol}$ and a high Eddington ratio $\lambda_{\rm Edd}=\frac{L_{\rm bol}}{L_{\rm Edd}}$ have always been excluded a priori as a potential host of CL events \citep{wang23}, recent studies proved that they can also be good candidates \citep{Miniutti13, Elitzur14, wang23, xu24}. From this point of view, narrow-line Seyfert 1 galaxies (NLS1s, \cite{Osterbrock85}) seem to be good candidates for hosting this CL phenomenon under those mechanisms which operate close to the Eddington limit \citep{xu24}. NLS1s are identified, by definition, with a Full Width at Half Maximum (FWHM) of H$\beta$ less than 2000 \kms \citep{Goodrich89} and an [O III]$\lambda$5007/H$\beta$ ratio of less than 3 \citep{Osterbrock85}. Many NLS1s also exhibit prominent Fe II multiplet emissions, often relatively stronger than the total H$\beta$ emission \citep{Osterbrock85, Rakshit17a}. Their black hole mass is believed to be lower ($<10^8$ M$_\odot$) than typical broad-line Seyfert 1 (BLS1, \citealp{Peterson11}). Since their bolometric luminosity is similar to that of BLS1s, they are probably characterised by an extremely high accretion rate \citep{Marziani14}, often near or even exceeding the Eddington limit \citep[Eddington ratio $ 0.1 \leq \varepsilon \leq 10 $][]{Boroson92, Komossa06, Jin17b, Tortosa22} .  This is likely due to accretion occurring in 'puffed-up' slim discs which are the ones able to sustain this extreme regime \citep{Wang03}, but also to block a significant part of the ionising continuum from reaching the NLR, bringing to the observed weak [O III] emission \citep[e.g.,][and references therein]{Luo15, Jin17a}. 

The disc instabilities produced in super-Eddington sources such as NLS1s are recognised as one of the physical mechanisms behind the CL AGN. In the X-ray band, NLS1s are known for their complex spectral features, including a steep X-ray slope ($\Gamma$ $\approx$ 2.0 $-$ 2.2, e.g., \citet{Boller00}) and an excess in the ultra-soft X-ray region (below 1 keV) (e.g., \citealp{Komossa00}), above the prediction of a single power law. The spectral shape and X-ray brightness are indeed further indicators of their high accretion rates. Their rapid and significant variability in the X-rays is another point in favour of NLS1s to be prone to CL events, since the great majority of these objects have been discovered due to peculiar behaviour in the X-ray \citep{Gallo18}.

In this paper we present an optical and X-ray study of a particular source, 2MASX J04130709-0050165, first classified as NLS1 in 2004, which has undergone X-ray flux variability and optical spectral type changes over a period of almost twenty years. In Sect.~\ref{sec:obs}, we describe the multiple observations we collect for this object, in Sect.~\ref{sec:opanalysis} and Sect.~\ref{sec:xanalysis} we present the analysis of the optical and X-ray spectral fitting. In Sect.~\ref{sec:disc} we discuss the results and give our conclusions in Sect.~\ref{sec:conclusion}. Additional technical details are given in Appendix.\\

For this analysis, we adopted a standard $\Lambda$CDM cosmology with a Hubble constant $H_{0} = 67.8$ km s$^{-1}$ Mpc$^{-1}$, considering a flat Universe with the matter density parameter $\Omega_{M} = 0.308$ and the vacuum density parameter $\Omega_{vac} = 0.692$ \citep{Planck16}.

%%%%%%%%%%%%%%%%%%%%%%%%%%%%%%%%%%%%%%%%%%%%%%%%%%%%%%%%%%%%%%%%%%%%%%%%%%%%%%%%%%%%%%%%%%%%%%%%%%%%%%%%%%%
\section{2MASX J04130709-0050165: multiple observations}
\label{sec:obs}

\begin{table}[]
    \centering
    \small
    \caption{Optical observations of J0413-0050.}
    \begin{tabular}{l c c c c c}
    \hline\hline
     Date &  Tel. &  Inst. & Exp. time (s) & S/N & D\\
     \hline\hline
      Apr 2004 & 6dF & 1200V & 1200 & 23.8 & 0.9\\
      Jan 2021 & NTT & EFOSC2 & 3600 & 24.8 & 0.99\\
      Dec 2021 & NOT & ALFOSC & 300 & 19 & 1.7\\
      Sep 2023 & VLT/UT1 & FORS2 & 300+300 & 25 & 0.75\\
      \hline
    \end{tabular}
    \tablefoot{Columns: (1) observation date; (2) telescope used; (3) instrument used. For 6df we directly reported the grism. For EFOSC2 spectrum we used the grism\#8, for ALFOSC we used grism\#7, for FORS spectrum we used grisms 600B+22 and 600RI+19; (4) exposure time in $s$, (5) S/N ratio in the 5100\AA{} continuum, (6) spectral dispersion in $\AA$pix$^{-1}$.}
    \label{tab:obs}
\end{table}

2MASX J04130709-0050165 (hereafter J0413-0050, R.A. 04h13m07.05s, Dec $-$00d50m16.63s, $z$=0.04) is an AGN which has been observed several times in the optical and X-ray range in the last two decades.\\
The observations are summarised in Table~\ref{tab:obs}. The first observation was performed in 2004 within the Six-degree Field Galaxy Survey (6dF) \citep{6dfsurveyplan} (hereafter 2004-04). It is worth noting that since the 6dF does not provide flux calibration, this spectrum was flux-calibrated by \citet{Chen18}, who created a calibration curve for the survey and analysed its optical properties. It must be said that the calibration function used by \citet{Chen18}, built as an average function for a sample of 167 NLS1s, may not be properly suitable for the spectrum of J0413-0050. However, \citet{Chen18} classified this source as NLS1 galaxy, measuring a FWHM(H$\beta$)= 2133 $\pm$ 634 \kms.\\
Meanwhile, the field of view (FoV) of this object was observed with the extended ROentgen Survey with an Imaging Telescope Array \citep[{\it eROSITA} hereafter,][]{2021A&A...647A...1P} in 2020 (ID: 065090-220/920).
Almost one year later, an optical spectrum of this object was taken with the European Southern Observatory (ESO) Faint Object Spectrograph and Camera v.2 (EFOSC2) mounted on the New Technology Telescope (NTT; proposal ID: NTT/106.21HS, PI M. Berton) in January 2021 (hereafter 2021-01). Noticing the drastic change between the 2004-04 and the 2021-01 spectra, another spectrum was taken with the Nordic Optical Telescope (NOT) at the end of 2021 (hereafter 2021-12).\\
One year later (November 2022), we asked for a Target of Opportunity (ToO) observation (ObsID 00015418001) with \textit{The Neil Gehrels Swift Observatory} \citep[\swift\ hereafter,][]{2004ApJ...611.1005G}. Considering the strong short-term X-ray variability for NLS1s \citep{Grupe04a}, we decided to ask for another \swift\ ToO in September 2023 observation (ObsID 00015418004).
To explore a possible relation between the X-ray flux variations and the variability seen in the optical range we asked for Director's Discretionary Time at ESO Unit Telescope 1 (UT1), right after this last \swift spectrum, requesting a FOcal Reducer/low dispersion Spectrograph 2 (FORS2) spectrum (hereafter 2023-09) covering the main emission lines for precise modelling of the line profile, searching for broad components to appear.\\
In conclusion, we collected four epochs of optical spectroscopy and three epochs of X-ray spectroscopy.

%%%%%%%%%%%%%%%%%%%%%%%%%%%%%%%%%%%%%%%%%%%%%%%%%%%%%%%%%%%%%%%%%%%%%%%%%%%%%%%%%%%%%%%%%%%%%%%%%%%%%%%%%%%
\section{Optical analysis: spectral fitting}
\label{sec:opanalysis} 

\subsection{Data preparation and fitting procedure}
\label{subsec:prep_fit}
Before applying line decomposition, all spectra were corrected for Galactic extinction with IRAF (version NOIRLab IRAF V2.18, \cite{Tody00}), adopting A(V)=0.38 and assuming a reddening law with R$_{v}$=3.1 \citep{Cardelli89}. The spectra were then shifted to the rest frame using \textit{z}=0.040178 (\href{https://ned.ipac.caltech.edu/byname}{NED}). The 2004-04 and 2021-12 spectra were already calibrated at this stage, whereas the 2021-01 and 2023-09 spectra were reduced using the ESO EFOSC and FORS2 pipelines through ESO Reflex \citep{Izzo10, Freudling13}.

To analyse and quantify the spectral variability of J0413$-$0050, we fitted its four optical spectra with the Fully Automated pythoN tool for AGN Spectra analYsis (\href{https://fantasy-agn.readthedocs.io/en/latest/index.html}{\texttt{fantasy}}, \cite{Ilic20, Rakic22, Ilic23}), a Python code for multi-component spectral fitting optimised for type 1~AGN, which simultaneously fits the underlying continuum and sets of emission lines on a wavelength range of 3700-11000 $\AA$.

To properly isolate the AGN contribution and, in particular, the NLR and BLR components of the emission lines, we removed the host galaxy contamination. The host galaxy subtraction was performed using a linear combination of quasar and galaxy eigenspectra \citep[for details see][]{Ilic23}, yielding a pure AGN spectrum in each case.

At this point, we applied the \texttt{fantasy} multi-component spectral fitting, from which we extracted their broad and narrow components to constrain the CL phenomenon. Within the 4000-7000 $\AA$ spectral range, we fit the model with a broken power-law continuum, broad lines (hydrogen Balmer lines as H$\alpha$, H$\beta$, H$\gamma$, and HeI, HeII) fixed to have the same width and shift; narrow lines (H$\alpha$, H$\beta$, H$\gamma$, [O III]$\lambda$4959,5007, [N II]$\lambda$6548,6583, SII$\lambda$6716,6731) tied to the [O III]$\lambda$5007 width and shift (the ratios of [O III] and [N II] doublets in particular were fixed to 3, \citealt{Osterbrocklibro, Dimitrijevic07, Dojvcinovic23}), and Fe II model in which all line have the same width and shift \citep[see][for details]{Ilic23}. The optical Fe II model, built on atomic data of the transitions \citep{Ilic23}, also includes the H$\alpha$ line region \citep[which is highly populated in NLS1s, see e.g.][]{Park22}, and that was not covered by the previous Fe II models developed in \cite{Kovacevic10, Shapovalova12}.

In the next subsection, we examine all the fitted spectra, describing the model and the parameters we applied.

\begin{figure*}[h!]
    \centering
    \includegraphics[width=18cm]{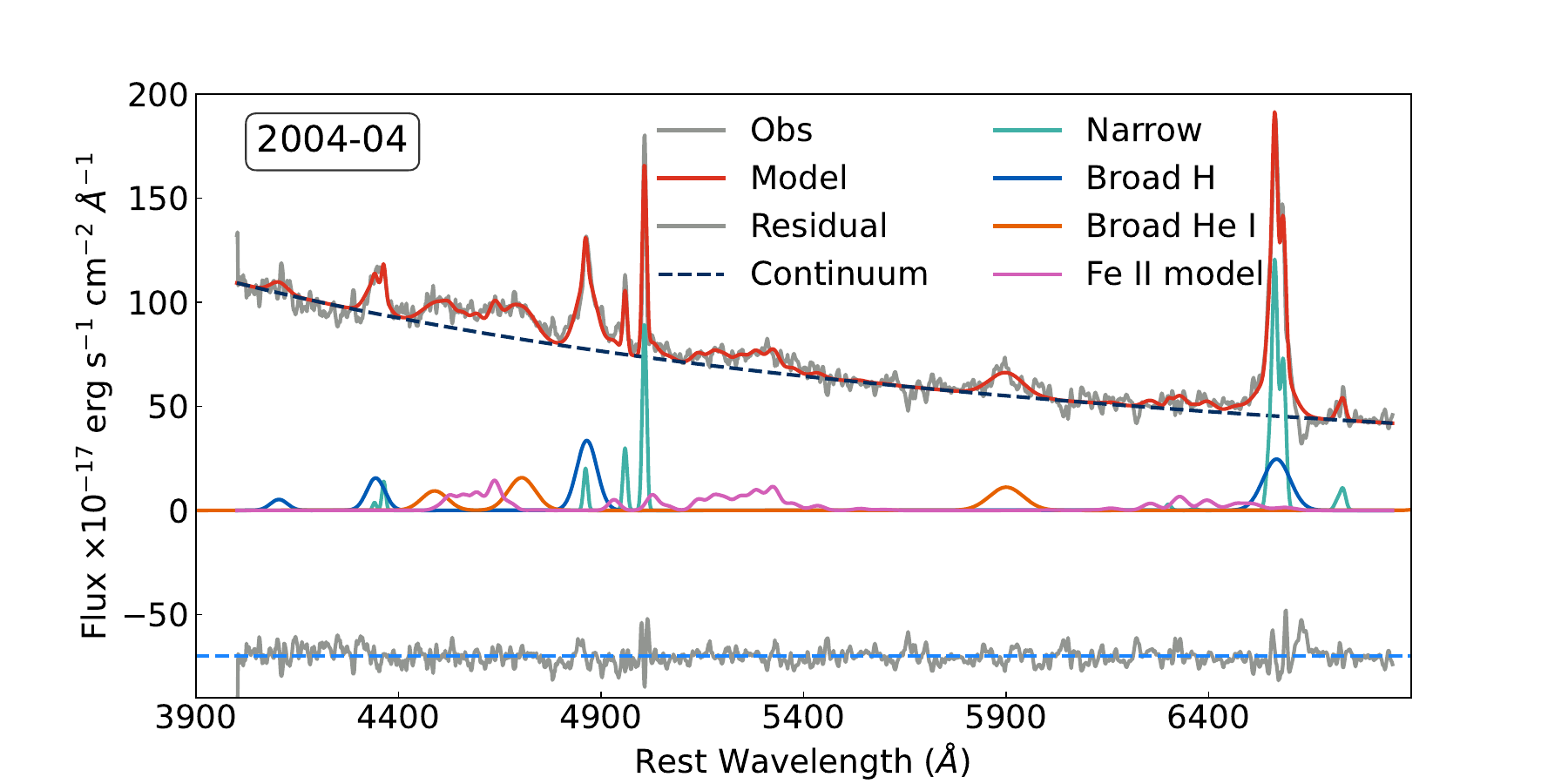}
    \caption{Multi-component fitting with \texttt{fantasy} of J0413-0050 2004-04 spectrum (grey line) in the 4000-6850 $\AA$ range. The model (red line) is composed of an underlying continuum (dark blue dotted line), broad components of Balmer lines (light blue line), narrow emission lines (water green line), Fe II multiplets (fuchsia line) and broad components of He I (orange). The residuals, subtraction of the model to the AGN spectrum, are shown in grey at the bottom of the panel.}
    \label{fig:6df}
\end{figure*}

%%%%%%%%%%%%%%%%%%%%%%%%%%%%%%%%%%%%%%%%%%%%%%%%%%%%%%%%%%%%%%%%%%%%%%%%%%%%%%%%%%%%%%%%%%%%%%%%%%%%%%%%%%%
\subsection{2004-04 spectrum}
\label{subsec:6df}
\renewcommand{\arraystretch}{1.5}
\begin{table*}[ht!]
\caption[]{Fitting parameters of the hydrogen Balmer lines}
\centering
\begin{tabular}{l l l l l l l}
\hline\hline
Spectrum   &  Flux(H$\alpha_{\rm b}$)  & Flux(H$\alpha_{\rm n}$) & Flux(H$\beta_{\rm b}$) & Flux(H$\beta_{\rm n}$) & FWHM$_{\rm b}$ &  FWHM$_{\rm n}$ \\ 
\hline\hline
2004-04 & 1896 $\pm$ 46 & 1596 $\pm$ 13 & 2108 $\pm$ 38 & 2656 $\pm$ 8 & 3675 $\pm$ 30 & 741 $\pm$ 3\\
2021-01 &  &  & & $<50 \pm 4$& & 350 $\pm$ 1\\
2021-12 & 700 $\pm$ 8 & 576 $\pm$ 7 & & & 2206 $\pm$ 50 & 374 $\pm$ 1\\
2023-09  & 579 $\pm$ 43 & 191 $\pm$ 6 & 119 $\pm$ 13 & 41 $\pm$ 4 & 1901 $\pm$ 44 & 284 $\pm$ 2\\ 
\hline 
\end{tabular}
\tablefoot{Columns: (1) date of the observation; (2,3,4,5) integrated flux of H$\alpha$ and H$\beta$ broad and narrow components in $10^{-17}$ erg s$^{-1}$ cm$^{-2}$; (6,7) FWHM of the broad and of the narrow components in \kms.}
\label{tab:balmerflux}
\end{table*}

\renewcommand{\arraystretch}{1.5}
\begin{table*}[ht!]
\caption[]{[O III] measurements}
\centering
\begin{tabular}{l l l l}
\hline\hline
Spectrum  &  Flux ([O III]$\lambda$4959) & FWHM & Flux ([O III]$\lambda$5007) \\ 
\hline\hline
2004-04  & 390 $\pm$ 3 & 741 $\pm$ 3 & 1069 $\pm$ 10\\
2021-01 & 454 $\pm$ 3 & 350 $\pm$ 1 & 1536 $\pm$ 10\\
2021-12& 443 $\pm$ 3 & 397 $\pm$ 1 & 1322 $\pm$ 9\\
2023-09 & 176 $\pm$ 2 & 283 $\pm$ 2 & 577 $\pm$ 7\\
\hline
\end{tabular}
\tablefoot{Columns: (1) date of the observation; (2) integrated flux of the [O III]$\lambda$4959 line in $10^{-17}$ erg s$^{-1}$ cm$^{-2}$; (3) FWHM in \kms of the narrow component; (4) integrated flux of the [O III]$\lambda$5007 line in $10^{-17}$ erg s$^{-1}$ cm$^{-2}$.}
\label{tab:oiii}
\end{table*}

\renewcommand{\arraystretch}{1.5}
\begin{table*}[ht!]
\caption[]{Luminosity measurements and Eddington ratio $\lambda_{\rm Edd}$}
\centering
\begin{tabular}{l l l l l}
\hline\hline
Spectrum  &  Flux (5100 $\AA$) & $\lambda$L (5100) &  L$_{\rm{bol}}$ & $\lambda_{\rm Edd}$ \\
& $10^{-17}$ erg s$^{-1}$ cm$^{-2}$ $\AA^{-1}$ & 10$^{43}$ erg s$^{-1}$ & 10$^{44}$ erg s$^{-1}$ &\\
\hline\hline
2004-04&  72 $\pm$ 2  &  1.47 & 1.32 & 0.46 \\
2021-01 &  23 $\pm$ 6  &  0.47 & 0.43 & 0.19 \\
2021-12&  20 $\pm$ 4  &  0.41 & 0.37 & 0.16 \\
2023-09 &  13 $\pm$ 1  &  0.27 & 0.25 & 0.12 \\ 
\hline
\end{tabular}
\tablefoot{Columns: (1) date of the observation; (2) mean density flux of the continuum at 5100 $\AA$ in $10^{-17}$ erg s$^{-1}$ cm$^{-2}$ $\AA^{-1}$; (3) luminosity of the continuum at 5100 $\AA$ in 10$^{43}$ erg s$^{-1}$; (4) bolometric luminosity in 10$^{44}$ erg s$^{-1}$; (5) Eddington ratio.}
\label{tab:lum}
\end{table*}

We initially smoothed the spectrum (each pixel was replaced with the average of 5 pixels) to reduce the noise and to facilitate comparison with the other spectra. Details on host galaxy subtraction will be described in the Sect.~\ref{subsec:6df_app}. Once we isolated the AGN spectrum, we fitted it with a multi-component spectral model, as described in the previous subsection. The results of the fitting are plotted in Fig.~\ref{fig:6df}.

The main Balmer lines, H$\alpha$ and H$\beta$, both show narrow and broad components, indicating an unobscured view of the central engine. The broad components resulted to have a FWHM of 3675 \kms while the narrow components of 741 \kms (Tab.~\ref{tab:balmerflux}). The integrated fluxes for the single components are reported in Tab.~\ref{tab:balmerflux}. Despite the H$\alpha$-[N II] complex not being completely resolved, due to poor spectral resolution, \texttt{fantasy} could decompose the H$\alpha$ contribution from the [N II] doublet. 

The flux measurements related to the [O III]$\lambda$4959,5007 lines in all four spectra are reported in Tab.~\ref{tab:oiii}. Comments on these results will follow in the Discussion section.
In 2004, J0413-0050 also showed strong Fe II multiplets, which are typical features of NLS1s. The total Fe II emission can be identified in three different wavelength bands (Fe II blue 4340-4680 $\AA$, Fe II green 5100-5600 $\AA$, Fe II red 6100-6650 $\AA$) as suggested by \citet{Ilic23} and, in our case, all of the three are identified. The Fe II multiplets can also be blended with the main emission lines, as in the case of [O III]$\lambda$5007, for which the Fe II components are responsible for a faint red wing. The Fe II red is also visible in this spectrum, but it does not affect the H$\alpha$-[N II] complex.

From the BLR radius and the line dispersion second-order moment, following the method in \citet{Berton15a}, \citet{Chen18} calculated a BH mass of (2.87 $\pm$ 0.79)$\times$10$^6$ M$_\odot$ for this source. We also measured the BLR radius from the relation found in \citet{Greene10}:

\begin{equation}
    \log \left( \frac{R_{\rm{BLR}}}{\mathrm{l.d.}} \right) = (1.85 \pm 0.08) + (0.53 \pm 0.03) \log \left( \frac{L(H\beta)}{10^{43}~\mathrm{erg~s^{-1}}} \right) \; ,
    \label{eq:Rblr_Hbeta}
\end{equation}

where $L(H\beta)$ is the integrated luminosity of the broad H$\beta$ component, and the BLR radius is expressed in light days. We obtained a value of about 6 light days for the $R_{\rm{BLR}}$. 
Taking as valid the mass measurements from \citet{Chen18}, we calculated the Eddington ratio as:

\begin{equation}
    \lambda_{\rm Edd} = \frac{L_{\rm{bol}}}{L_{\rm{Edd}}} = \frac{L_{\rm{bol}}}{1.3\times10^{38}M_{\rm{BH}}/\rm M_{\odot}} \; ,
    \label{eq:Edd}
\end{equation}

where $L_{\rm{bol}}$ is the bolometric luminosity and $L_{\rm{Edd}}$ is the Eddington luminosity \citep{BeckmannShrader}. The Eddington luminosity is $L_{\rm{Edd}}$ = 3.73 $\times$ 10$^{44}$ erg s$^{-1}$.
%Check from here
The bolometric luminosity may be estimated from the continuum luminosity at 5100$\AA$, although bolometric corrections may depend on luminosity and accretion rate \citep[e.g.,][]{Runnoe12, Duras20}, typically with relatively small uncertainties. Since no bolometric calibration has yet been specifically tested on systems undergoing strong accretion‑state variations, we adopt the prescription of \citet{Netzer19}. This relation is calibrated on a large sample spanning a wide luminosity range, including high‑Eddington sources (up to $\lambda_{\rm Edd} \approx 0.5$ for thin accretion discs), and has been employed in recent studies of variability and CL AGN \citep[e.g.,][]{Bing25, Kollatschny26}.
The \citet{Netzer19} formula represents a revised version of the classical relation  ($L_{\rm bol}$=9$\lambda$$L_{\lambda}$5100, \citealp{Kaspi00}), in which the bolometric correction factor $k_{\rm bol}$ explicitly depends on inclination:

\begin{equation}
    L_{\rm{bol}} = k_{\rm bol} \lambda L_{\lambda}(5100\AA); \quad k_{\rm bol} = 40 \times \left[\frac{\lambda L_{\lambda}(5100\AA)}{10^{42} \rm erg/s}\right]^{-0.2} \times \left[\frac{1}{f_i}\right].
    \label{eq:Edd}
\end{equation}

For typical Type~1 AGN, which generally have inclinations ($i$) around 56$^\circ$, the corresponding correction factor ($f_i$) is approximately 1.4, while for nearly face‑on accretion discs it increases to about 2.5 \citep{Netzer19}. Since the inclination of our source cannot be constrained, we adopt an average $f_{i}=$2, following the approach of \citet{Crepaldi25, DallaBarba26}.
This relation can be affected by the host galaxy component \citep{Crepaldi25}. For this reason, we measured the flux of the continuum at $\lambda$=5100$\AA$ on the spectrum for which the host contribution had already been subtracted, and then we retrieved the relative luminosity. These values are reported in Tab.~\ref{tab:lum}.

%%%%%%%%%%%%%%%%%%%%%%%%%%%%%%%%%%%%%%%%%%%%%%%%%%%%%%%%%%%%%%%%%%%%%%%%%%%%%%%%%%%%%%%%%%%%%%%%%%%%%%%%%%%
\subsection{2021-01 spectrum}
\label{subsec:NTT}

\begin{figure*}[h!]
    \centering
    \includegraphics[width=18cm]{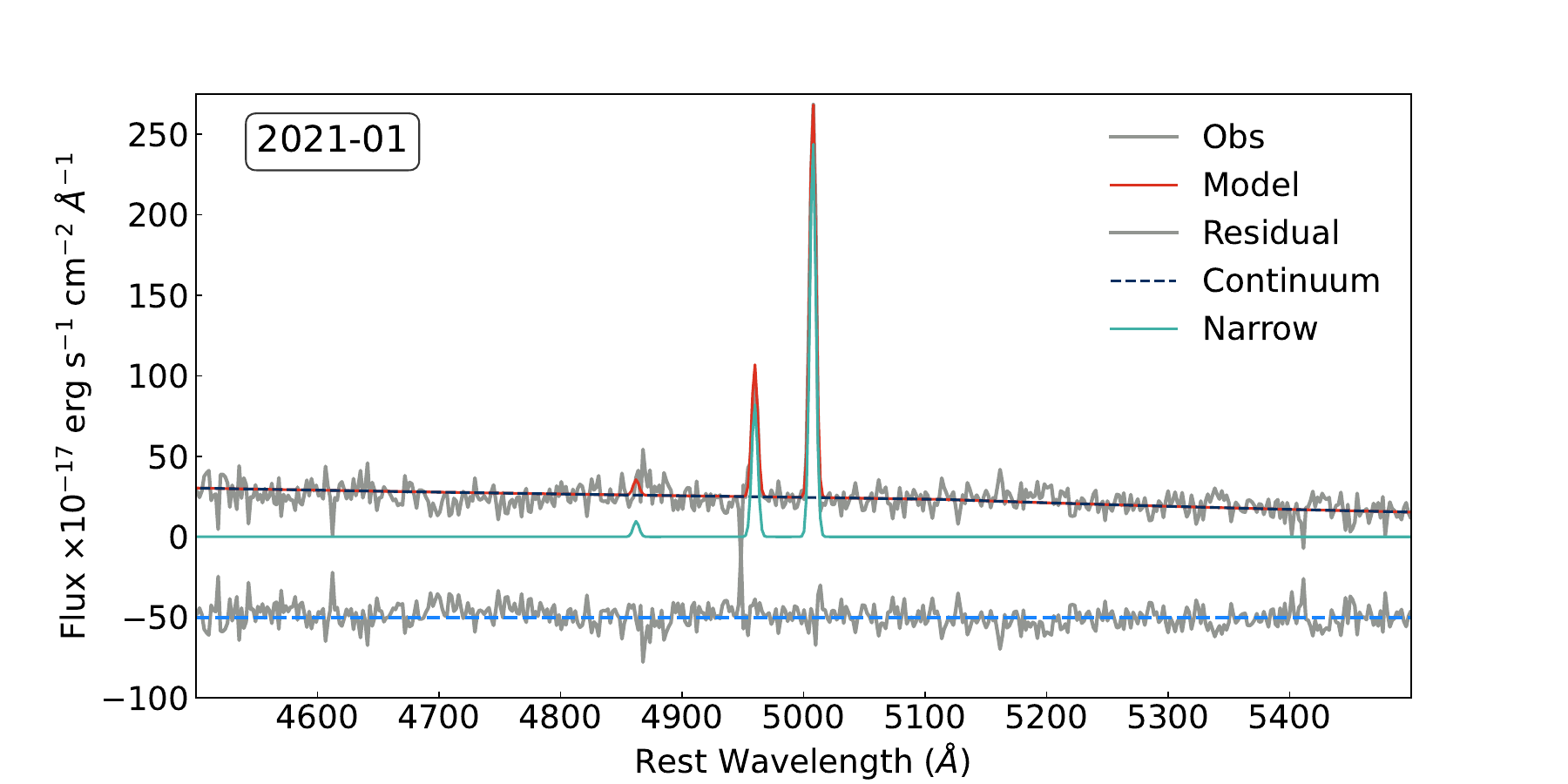}
    \caption{Multi-component fitting with \texttt{fantasy} of J0413-0050 2021-01 spectrum (grey line) in the 4300-5700 $\AA$ range. The model (red line) is composed of an underlying continuum (dark blue dotted line)and narrow emission lines (water green line). The residuals, subtraction of the model to the AGN spectrum, are shown in grey at the bottom of the panel.}
    \label{fig:NTT1}
\end{figure*}

The host galaxy modelling and subtraction is described and shown in Sect.~\ref{subsec:NTT_app}. Also in this case, we obtained a rising continuum towards the blue wavelength for the pure AGN spectrum, with a steep slope. We then built up the model as we did for the 2004-04 spectrum and the result of the fitting is plotted in Fig.~\ref{fig:NTT1}.

Although H$\beta$ was identified in the fit, both its broad and narrow component show an amplitude lower than 14.55, the value of the root-mean-square (rms or $\sigma$) measured in an interval of 100 pix around 5100 $\AA$. The same issue is seen for the Fe II multiplets amplitude, which stays under the 2$\sigma$. For this reason, it is a reliable hypothesis that those are not real lines, rather \texttt{fantasy} fits the noise. The H$\alpha$ is instead not present in this spectral range.  We can affirm that emission lines coming from the BLR are not present in these spectra.
 
Regarding the oxygen lines, their fluxes are consistent with measurements done on the 2004-04 spectrum (Tab.~\ref{tab:oiii},~\ref{tab:balmerflux}). The computed values for the several luminosities and Eddington ratio are listed in Tab.~\ref{tab:lum}.

%%%%%%%%%%%%%%%%%%%%%%%%%%%%%%%%%%%%%%%%%%%%%%%%%%%%%%%%%%%%%%%%%%%%%%%%%%%%%%%%%%%%%%%%%%%%%%%%%%%%%%%%
\subsection{2021-12 spectrum}
\label{subsec:NOT} 

\begin{figure*}[h!]
    \centering
    \includegraphics[width=18cm]{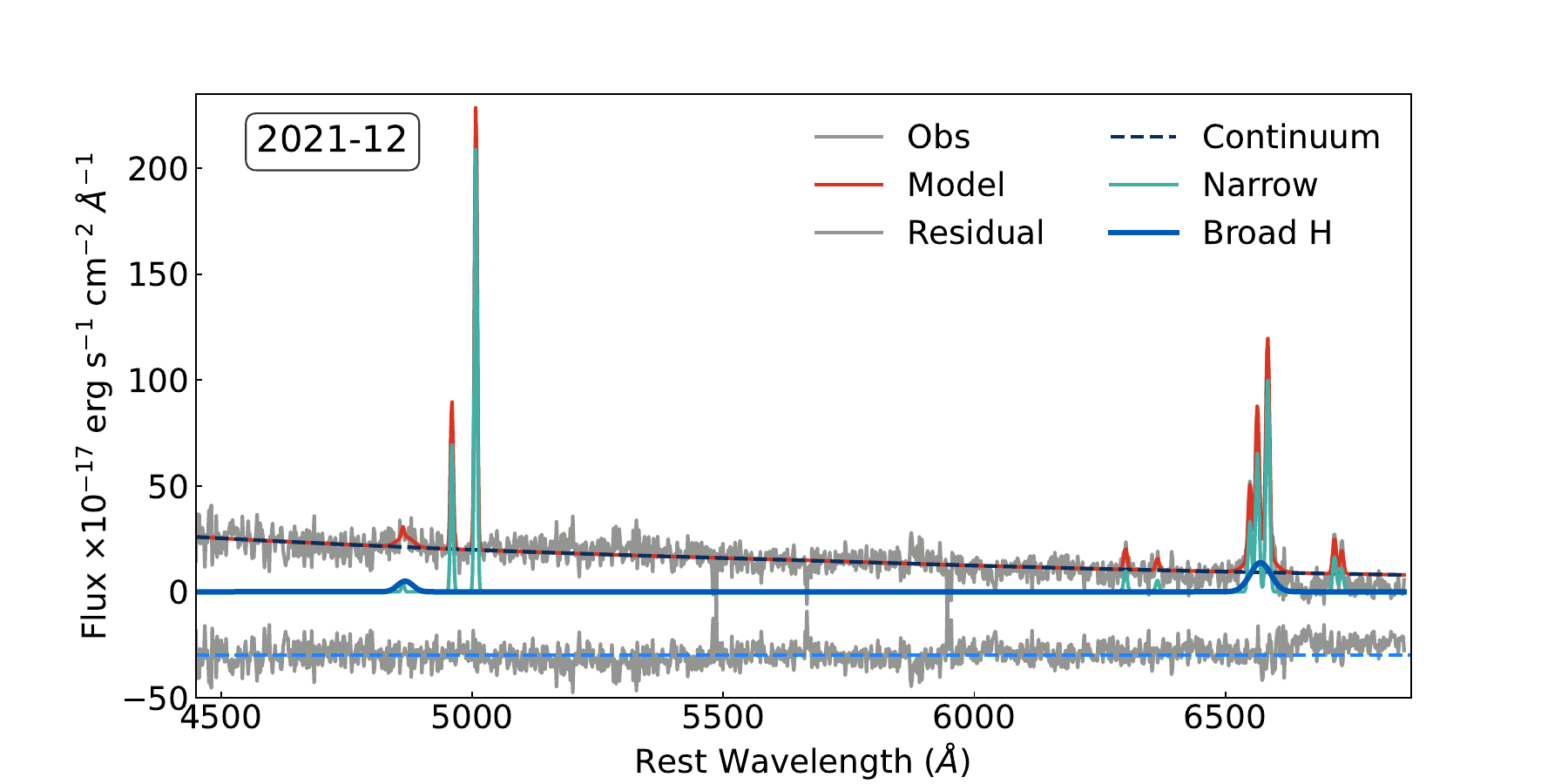}
    \caption{Multi-component fitting with \texttt{fantasy} of J0413-0050 2021-12 spectrum (grey line) in the 4500-6850 $\AA$ range. The model (red line) is composed of an underlying continuum (dark blue dotted line) and narrow emission lines (water green line) and broad emission lines (light blue line). No relevant broad components for H$\beta$ and He or FeII multiplets are present in this fit. The residuals, subtraction of the model to the AGN spectrum, are shown in grey at the bottom of the panel.}
    \label{fig:NOT_fit}
\end{figure*}

\begin{figure}[h]
\centering
\begin{minipage}{\linewidth}
    \centering
    \includegraphics[width=\linewidth]{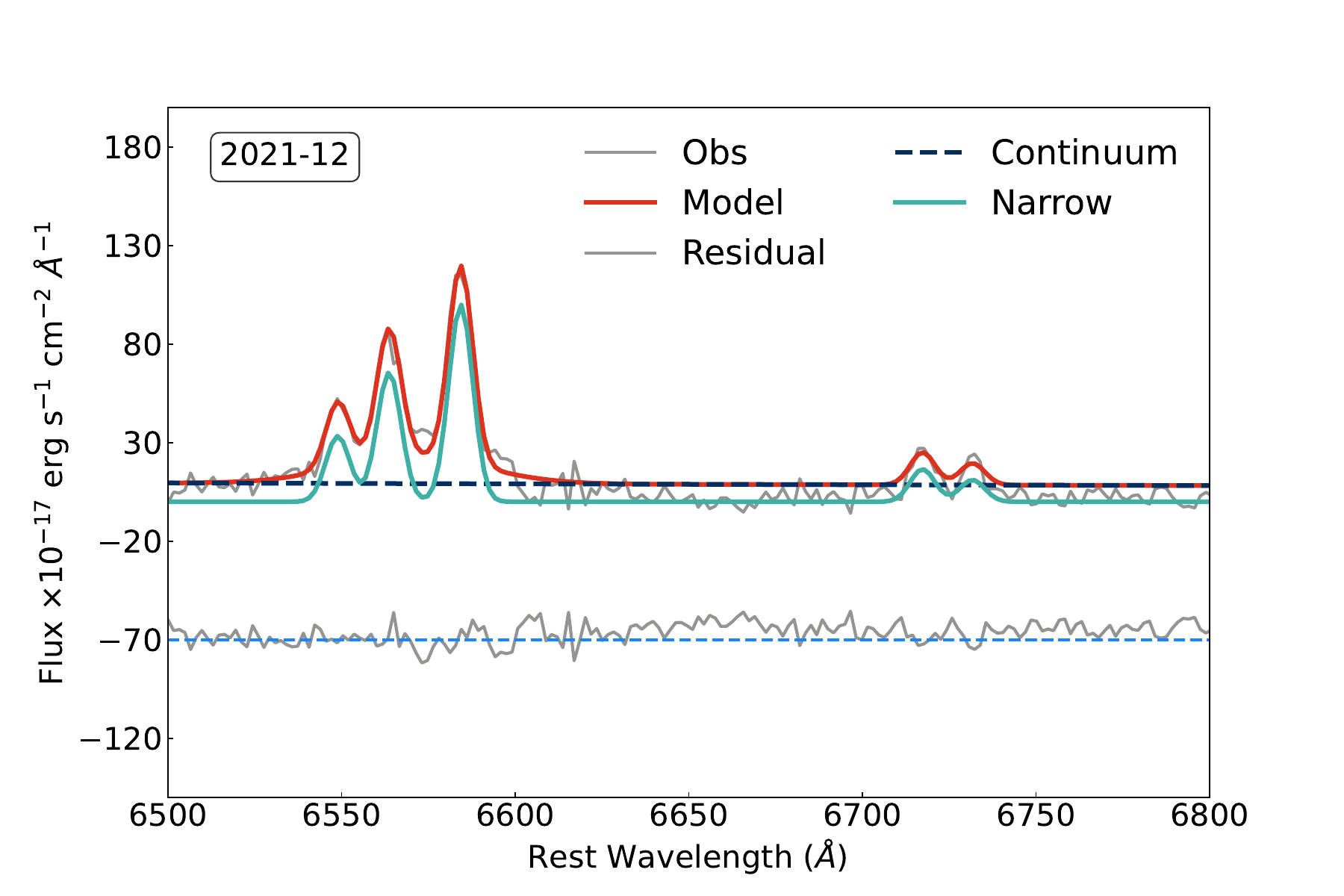}
    \caption{Zoom-in of the H$\alpha$-[N II] complex for the 2021-12 spectrum, which is totally resolved. A first attempt of fitting only narrow components of [N II] and H$\alpha$.}
    \label{fig:NOT_halpha}
\end{minipage}
\hfill
\begin{minipage}{\linewidth}
    \centering
    \includegraphics[width=\linewidth]{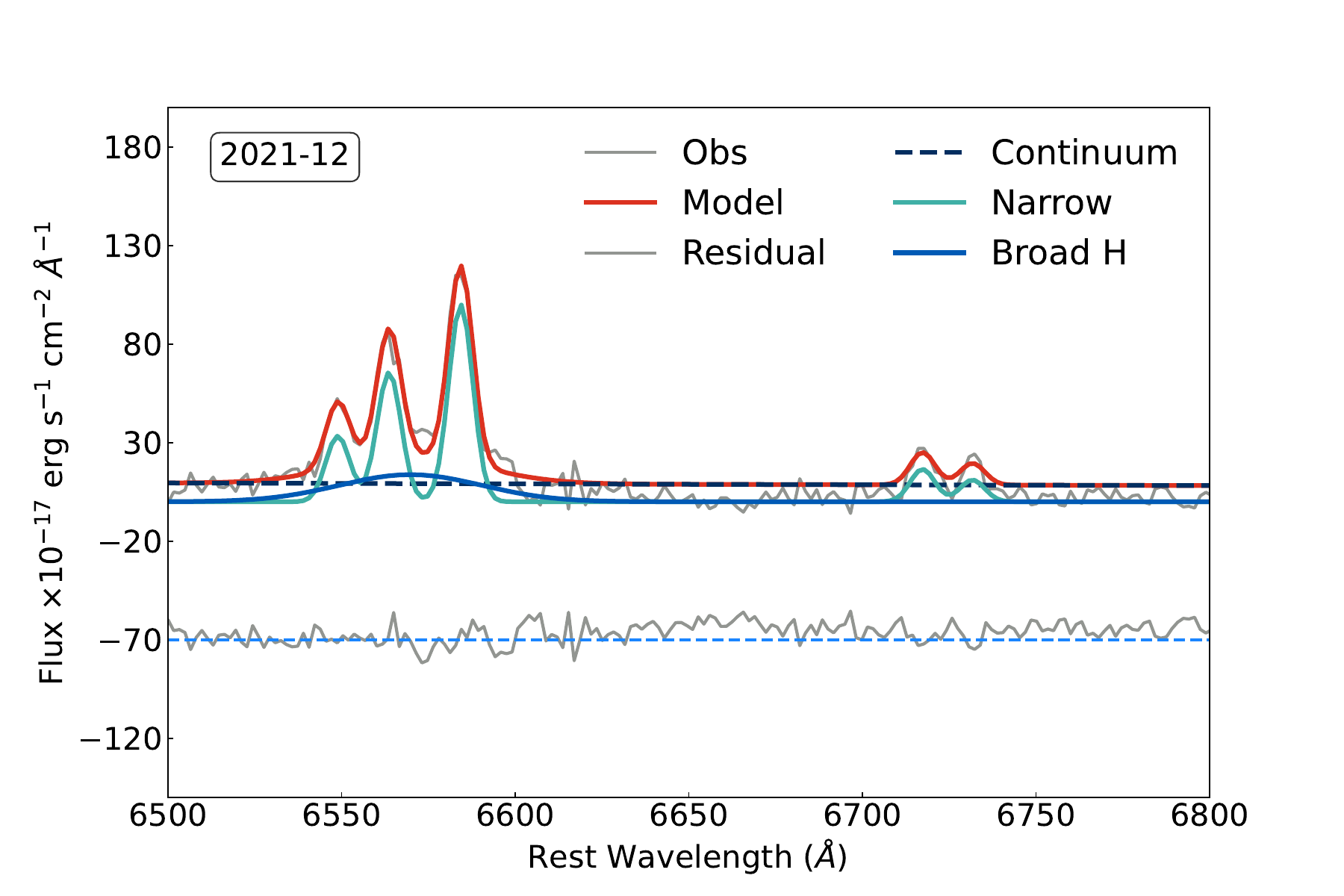}
    \caption{Zoom-in of the H$\alpha$-[N II] complex of the 2021-12 spectrum, including also the broad component for H$\alpha$, which amplitude results to be slightly higher than 3$\sigma$.}  
    \label{fig:NOT_halphab}
\end{minipage}
\end{figure}

\begin{figure}[h!]
    \centering
    \includegraphics[width=9cm]{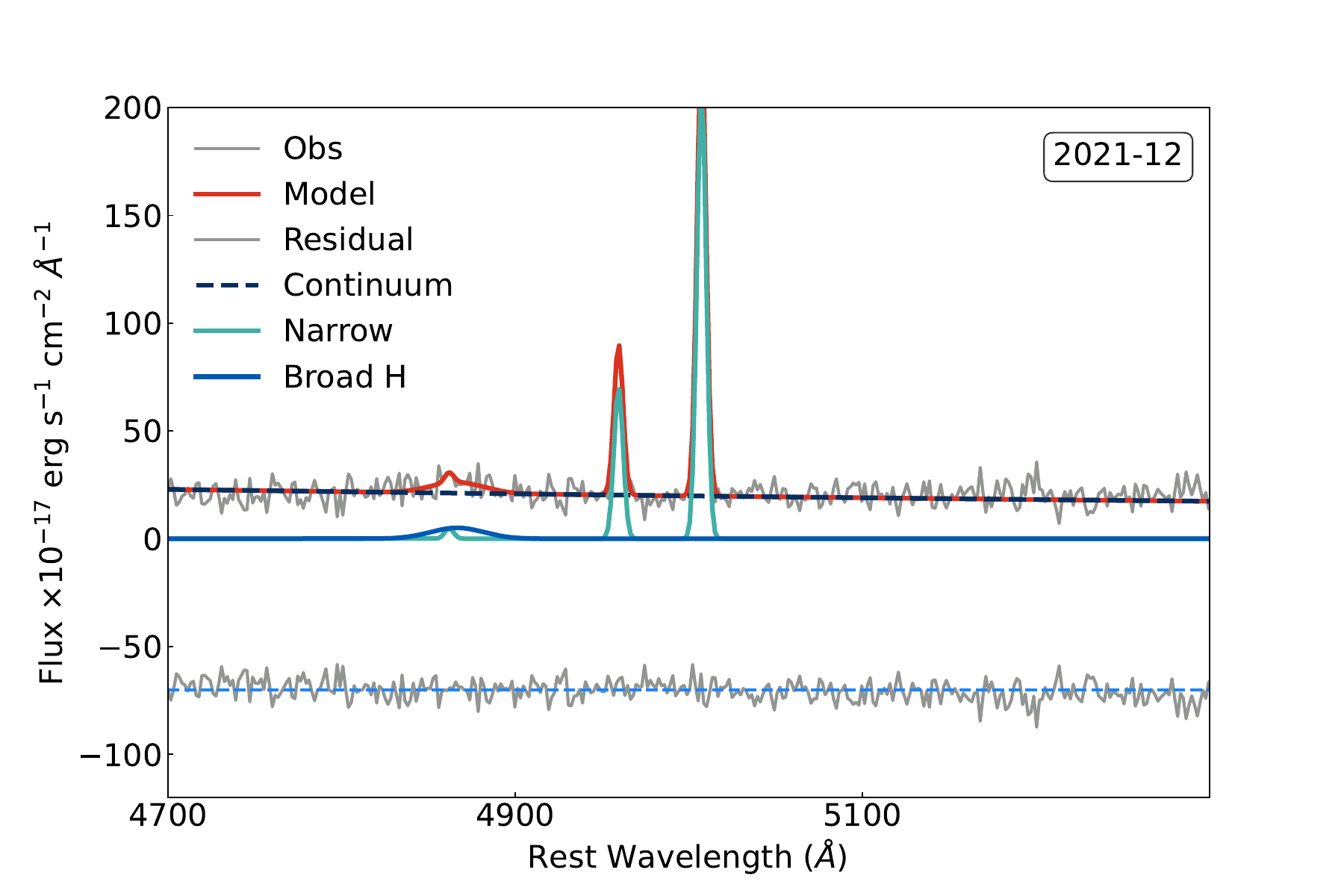}
    \caption{Zoom‑in of the H$\beta$–[OIII] region of the 2021‑12 spectrum, showing the attempted fit of an H$\beta$ component despite the absence of a real emission line at that position.}
    \label{fig:NOT_hbeta}
\end{figure}

The host galaxy model we retrieved from this spectrum is described and shown in Sect.~\ref{subsec:NOT_app}. The continuum slope of the pure AGN spectrum we obtained is rising towards the blue wavelengths. As for the previous observations, we fitted the 2021-12 spectrum with a multi-component spectral model. Although we initially included the He lines and the Fe II multiplets, their amplitude turned out to be negligible, and we therefore excluded them from the model. 

The result of the fit are shown in Fig.~\ref{fig:NOT_fit}. Within the spectral range of 4500-6850 $\AA$ only the [O III]$\lambda$4959,5007 lines and the resolved H$\alpha$-[N II] complex (zoom-in in Fig.~\ref{fig:NOT_halpha}) were detected. In a second attempt, we also fitted the H$\alpha$-[N II] complex including the broad component of H$\alpha$ (zoom-in in Fig.~\ref{fig:NOT_halphab}), whose amplitude resulted to be exactly at 3$\sigma$ (the rms in this spectrum is 4.592). The FWHM of both components are reported in Tab.~\ref{tab:balmerflux}. The two different models, with and without broad components, yield similar value for the $\chi^{2}$, 35 and 34.3, respectively. Conversely, the peaks of the narrow and broad line fitted in the spectra at the H$\beta$ line position lie at the same level of the rms indicating that this component is not significantly detected and can be attributed to the noise (zoom-in in Fig.~\ref{fig:NOT_hbeta}). We can confirm that throughout the entire year 2021, the H$\beta$ emission line completely disappeared from the spectra of J0413-0050, while the H$\alpha$ broad component may still be present in this spectrum. This would correspond to an IS~1.9 classification (broad component is only visible in H$\alpha$)although the H$\beta$ narrow component is absent.

The oxygen lines are comparable to the ones shown in the 2021-01 spectrum, although being slightly fainter (Tab.~\ref{tab:oiii}). Finally, we computed the luminosities (Tab.~\ref{tab:lum}), obtaining lower values than those measured in the earlier spectra.

%%%%%%%%%%%%%%%%%%%%%%%%%%%%%%%%%%%%%%%%%%%%%%%%%%%%%%%%%%%%%%%%%%%%%%%%%%%%%%%%%%%%%%%%%%%%%%%%%%%%%%%%%%%
\subsection{2023-09 spectrum}
\label{subsec:UT1}

\begin{figure*}[h!]
    \centering
    \includegraphics[width=18cm]{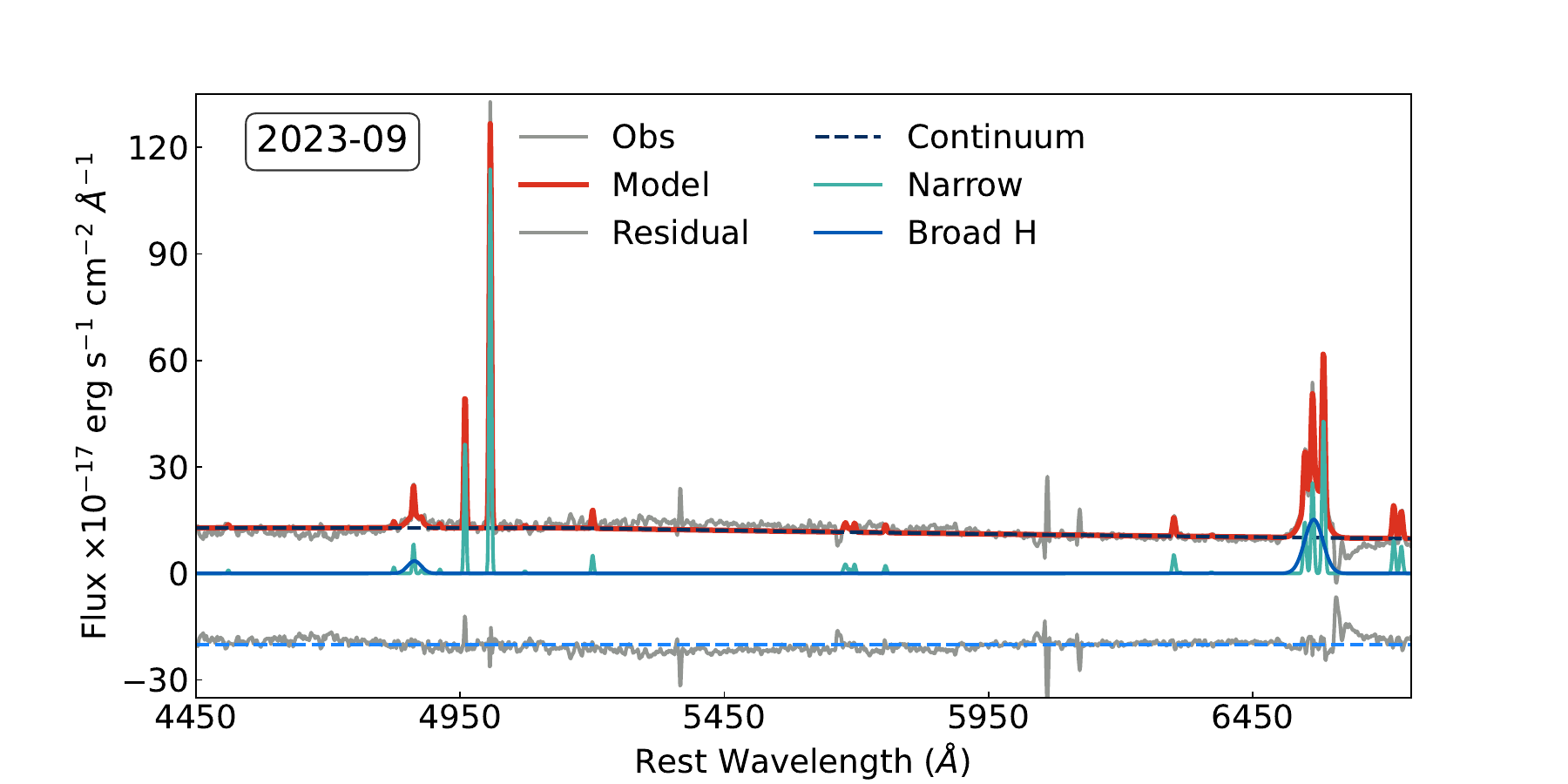}
    \caption{Multi-component fitting with \texttt{fantasy} of J0413-0050 2023-09 spectrum (grey line) in the 4450-6740 $\AA$ range. The model (red line) is composed of an underlying continuum (dark blue dotted line), broad components of Balmer lines (light blue line) and narrow emission lines (water green line). The residuals, subtraction of the model to the AGN spectrum, are shown in grey at the bottom of the panel.}
    \label{fig:UT1f}
\end{figure*}

\begin{figure}
    \centering
    \includegraphics[width=10cm]{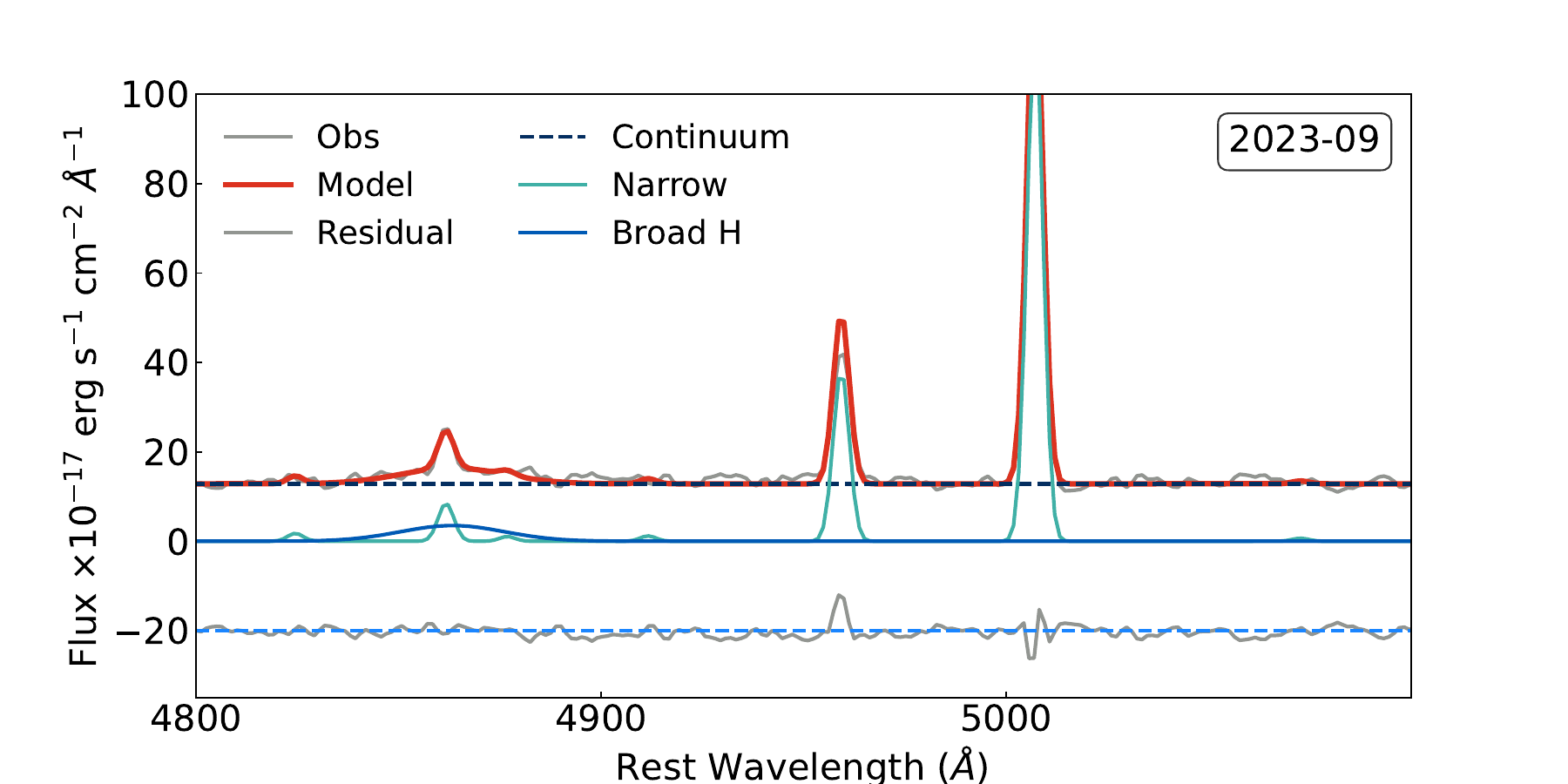}
    \caption{Zoom-in of the H$\beta$-[O III]$\lambda$4959,5007 region of the 2023-09 spectrum. H$\beta$ is decomposed in a broad and a narrow component, fixed to the [O III]$\lambda$5007 line. We cut the plot in flux to zoom into the H$\beta$ component, keeping out the majority of the emission of the [O III]$\lambda$4959,5007 lines.}
    \label{fig:UT1_hbeta}
\end{figure}

\begin{figure}
    \centering
    \includegraphics[width=10cm]{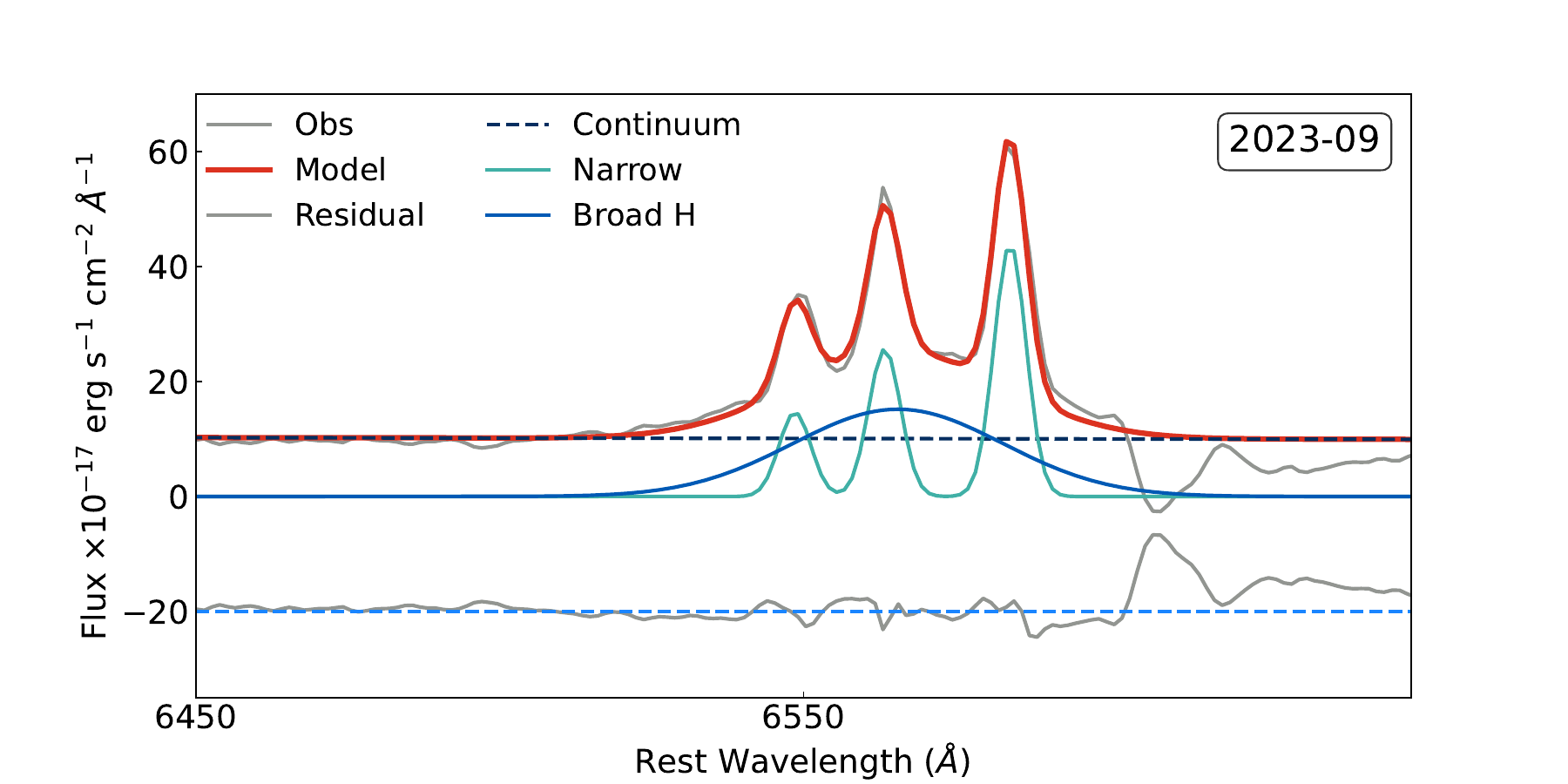}
    \caption{Zoom-in of the H$\alpha$-[N II] complex of the 2023-09 spectrum, which is  completely resolved. The  H$\alpha$ clearly show both broad and narrow components.}
    \label{fig:UT1_halpha}
\end{figure}

Once the AGN spectrum was isolated (Fig.~\ref{fig:UT1_host}), we fitted it with a multi-component spectral model, the same as for the 2004-04 spectrum. The results of the multi-component fitting are shown in Fig.~\ref{fig:UT1f}.

In the spectral range 4450-6740 $\AA$, \texttt{fantasy} identified both broad and narrow components for the Balmer lines (H$\alpha$, H$\beta$ and H$\gamma$), Fe II multiplets in the blue and green wavelength range, faint He emission and forbidden narrow lines. In this spectrum of September 2023, emission lines coming from the BLR (broad components of the Balmer lines, FeII multiplets) reappeared in the spectrum, indicating a clear view of the central region of the AGN. The H$\alpha$-[N II] complex is completely resolved, showing both narrow and broad components (Fig.~\ref{fig:UT1_halpha}) such as H$\beta$ (Fig.~\ref{fig:UT1_hbeta}). The FWHM of the broad and the narrow components are and their integrated fluxes are reported in Tab.~\ref{tab:balmerflux}. Those values seem to point again towards an IS classification,  which can also be determined through the ratio between the [O III]$\lambda$5007 and the H$\beta$ broad component flux \citep[R,][]{Whittle92}. In this spectrum, the object showed an R$\approx$4.8, which brings to IS~1.8 classification.

The oxygen fluxes peak at half the value reported in the 2021-01 and 2021-12 spectra (Tab.~\ref{tab:oiii}) while the continuum level is the same as for the 2021-12 and 2021-01 spectra (Fig.~\ref{fig:comparison}). Since these three spectra have been taken with the same slit aperture and seeing conditions, this oxygen flux decrease could be due to the different position angle (PA) of the slit for the three observations (Fig.~\ref{fig:pa}). Also for this spectrum, we computed the values for the luminosities (Tab.~\ref{tab:lum}).

%%%%%%%%%%%%%%%%%%%%%%%%%%%%%%%%%%%%%%%%%%%%%%%%%%%%%%%%%%%%%%%%%%%%%%%%%%%%%%%%%%%%%%%%%%%%%%%%%%%%%%%%%%%
\section{X-ray analysis}
\label{sec:xanalysis}

\begin{figure}[h!]
    \centering
    \includegraphics[width=9cm]{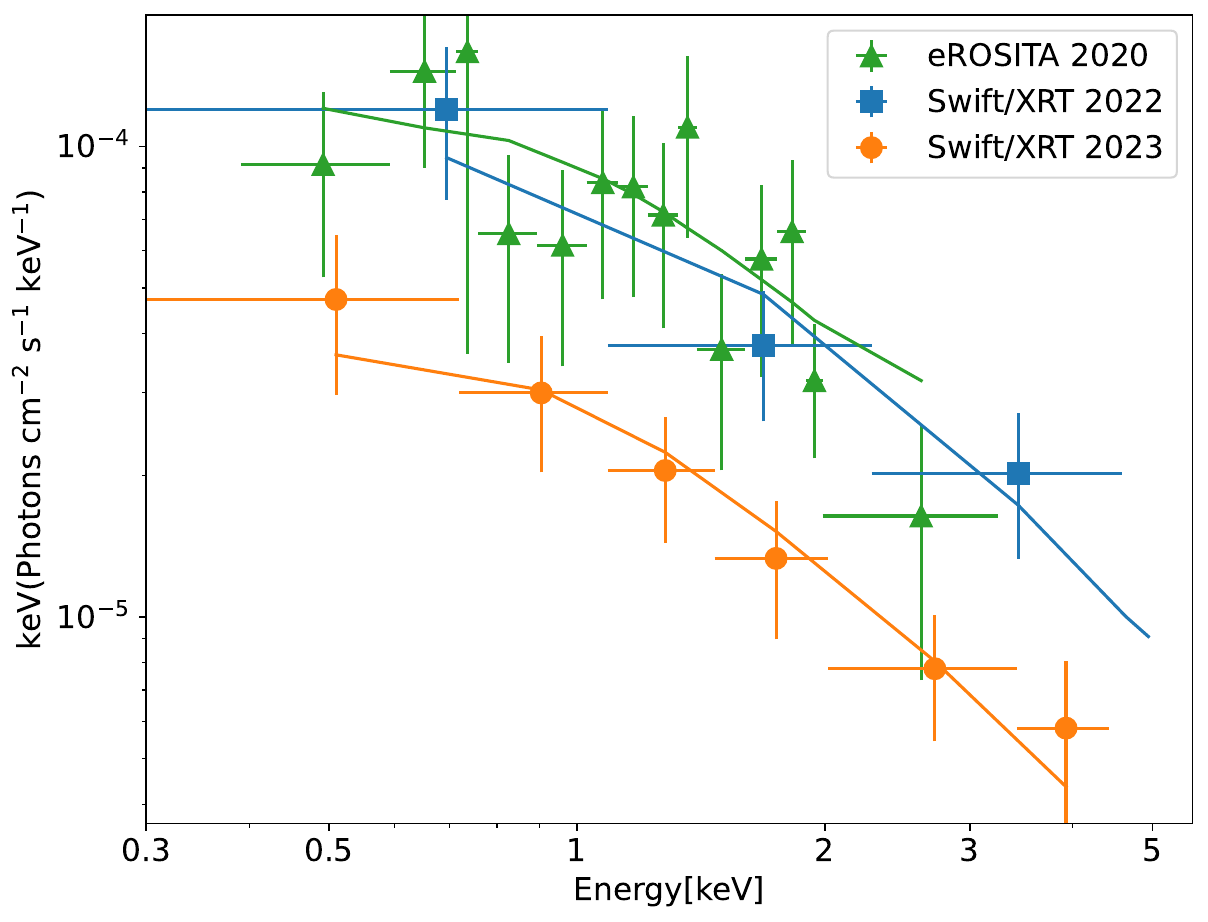}
    \caption{X-ray spectra of J0413-0050 from the {\it eROSITA} observation from 2020 (green triangles), {\it Swift}-XRT observation from 2022-11 (blue squares) and 2023-09 (orange circles). X-ray spectra are fitted with a simple power-law absorbed by Galactic absorption. Spectral are visually rebinned for clarity. Spectra above 5 keV are dominated by noise.}
    \label{fig:xplot}
\end{figure}

\begin{table*}[ht!]
\caption[]{Details of the X-ray observations analysed in this work together with X-ray flux, luminosity, $\Gamma$ measurements and Eddington ratio.}
\centering
\begin{tabular}{l l l l l l l l} 
X-ray telescope  & OBSID & Exp. time &  count rate & Flux$_{2-10\,keV}$ & L$_{2-10\,keV}$ & $\Gamma$ & $\lambda_{\rm Edd}$ \\ 
\hline\hline
{\textit{eROSITA}} 2020 & 065090-220/920 & $\sim 2.7$\,ks tot & $ 4.36\times10^{-2}$ & $6.69\times10^{-13}$ & $2.5\times10^{42}$ &$1.52\pm0.42$ & 0.11 \\
\hline
{\it Swift}/XRT 2022-11 & 00015418001 & $\sim 3.1$\,ks & $9.99\times10^{-3}\,$ &$5.73\times10^{-13}$& $2.11\times10^{42}$ & $1.58\pm0.53$ & 0.09 \\
\hline
{\it Swift}/XRT 2023-09 & 00015418002/4 & $\sim 16.0$\,ks tot &$3.73\times10^{-3}\,$ & $1.88\times10^{-13}$& $7\times10^{41}$  & $1.63\pm0.40$ & 0.03 \\
\hline
\end{tabular}
\tablefoot{The count rates refer to the 0.5--3~keV band for the \textit{eROSITA} observations and to the 0.5--5~keV for the {\it Swift}/XRT observations; 2--10~keV Flux in $erg/\rm cm^2/\rm s$ corrected for Galactic absorption; 2--10~keV luminosity in erg$/\rm s$.}
\label{tab:xray}
\end{table*}

J0413-0050 was also observed in the X-rays in 2020 by the {\it eROSITA}, in 2022 and 2023 with {\it Swift}/XRT. The {\it Swift}/XRT data were reduced following standard procedures using the latest calibration files. Details of the observations are reported in Table\,\ref{tab:xray}. 
{\it Swift}/XRT spectra were extracted using the \textsc{xselect} line interface (v2.4k) within the \textsc{heasoft} package (version 6.28). The background extractions are measured in an annular region with radius from $30\arcsec$to $60\arcsec$. If there is pile-up, the measured rate of the source is high (above $\sim$0.6 counts\,s$^{-1}$ in the Photon-Counting Mode). The {\it Swift}/XRT spectra of the sources of our sample resulted to not have a high pile-up degree; the source's extraction regions are measured using a circular region with a radius of $20\arcsec$. 

The {\it eROSITA} data were retrieved from the public archive\footnote{\url{https://erosita.mpe.mpg.de/dr1/erodat/}} as processed spectral products (including source and background spectra and response files), and thus no additional data reduction was required prior to spectral analysis.

The spectral analysis has been performed with the \textsc{xspec} v.12.11.1b software package \citep{Arnaud1996}. We performed the modelling by using the Cash statistics with direct background subtraction \citep[W-stat in \texttt{xspec},][]{1979ApJ...228..939C,1979ApJ...230..274W}. Due to the low net counts, a simple model composed by a primary power-law absorbed by the Galactic column density at the position of the source ($N_{\rm H}$=$1.25\times10^{21}\, cm^{-2}\rm$, \citealp{HI4PI2016}), was employed (\textsc{xspec} model: Tbabs * (powerlaw)).

The 2--10\,keV fluxes of J0413-0050 are reported in Tab.~\ref{tab:xray}, which also shows the $\Gamma$ values and the count rate. X-ray flux between the two {\it Swift}/XRT observations drops by a factor $\sim 3$ (i.e. $\sim0.5$-dex, see Fig.~\ref{fig:xplot}). Due to the low net count rate of the observations, constraining the intrinsic column density, $N_{\rm H}$, required to determine the level of obscuration, and any relative changes, was not possible. Despite this variability, the photon index remains consistent within uncertainties, indicating no significant spectral evolution. This suggests that the observed variability is primarily driven by changes in the normalisation of the primary continuum rather than by strong modifications of the coronal properties or variations in the line-of-sight absorption. No clear evidence for a “softer-when-brighter” trend is observed.

We also estimated the Eddington ratio by deriving the bolometric luminosity from the 2--10\,keV luminosity, following the relation presented in \citet{Gupta24}:
\begin{equation}
    L_{\rm{bol}} = k_{\rm bol} \times L_{2-10\,keV}; \quad k_{\rm bol} = 15.86 .
    \label{eq:Edd1}
\end{equation} 

This calibration accounts for the known dependence of X‑ray bolometric corrections on luminosity and Eddington ratio, as discussed in \citet{Gupta24}, and it is based on a large and homogeneous sample of unobscured AGN with simultaneous optical-to-X-ray observations. The resulting Eddington ratios (Tab.~\ref{tab:xray}) are consistent with the estimates derived from the optical spectra.

%%%%%%%%%%%%%%%%%%%%%%%%%%%%%%%%%%%%%%%%%%%%%%%%%%%%%%%%%%%%%%%%%%%%%%%%%%%%%%%%%%%%%%%%%%%%%%%%%%%%%%%%%%%
\section{Discussion}
\label{sec:disc}

\begin{figure*}
    \centering
    \includegraphics[width=18cm]{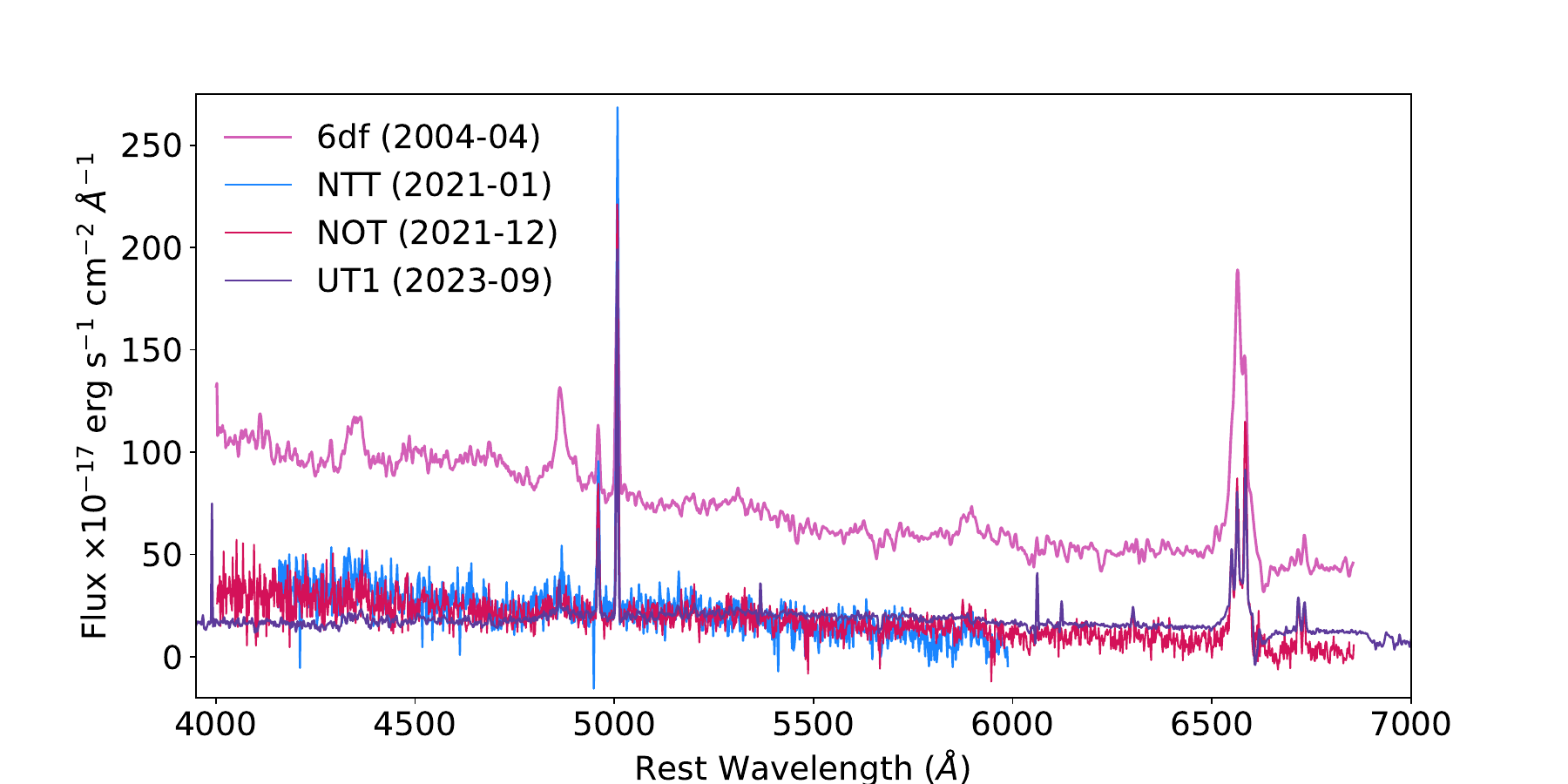}
    \caption{All the host subtracted spectra of J0413-0050. The H$\beta$ line is only visible in 2004-04 and 2023-09 spectra, while the spike at same position in the 2021-01 one is probably due to the noise. It is also shown the change in the H$\alpha$ and the continuum shapes.}
    \label{fig:comparison}  
\end{figure*}

\subsection{First phase: 2004 - 2022}
During the last 20 years, J0413-0050 has changed its optical classification, switching from an NLS1s (2004-04) to a IS~1.8 (2023-09), passing through a high-Eddington phase without showing any broad component for the Balmer lines (2021-01) and through a IS~1.9-like phase (2021-12), both lacking any evidence of the H$\beta$ line. Figure~\ref{fig:comparison} shows all the host-subtracted spectra of J0413-0050, highlighting the different spectral features at different epochs, such as continuum levels and shape, presence or absence of the Balmer lines, and the width and amplitude of the emission lines. The appearance and disappearance of the broad lines in the optical spectra, leading to changes in the spectral type, suggest that this source is a CS-AGN. The X-ray flux remained almost constant within the observations taken in 2020, 2022 and 2023, showing a stronger but mild decrease between the last two spectra. Since it was not possible to retrieve the N$_{\rm H}$ values, we cannot determine whether the source experienced an additional CO-AGN phase between 2014 and 2022.

The 2004-04 optical spectrum (Fig.~\ref{fig:6df}) showed both broad and narrow components of the main Balmer lines and a continuum rising towards the blue, typical of Type~1 AGN. At that time, the source was classified as an NLS1. However, as already noted, the flux calibration may have misestimated the continuum level. At the beginning of 2021, the spectrum showed the complete disappearance of the H$\beta$ line, while nothing could be said on the H$\alpha$ line, which was not included in the wavelength range (Fig.~\ref{fig:NTT1}).

This first change between two optical spectral types can be interpreted as a transition phase occurring at some point between 2004 and 2021, possibly due to the 'switching-off' of the central engine, followed by a 'switching-on' phase which could have begun shortly before the 2021-01 spectrum, potentially explaining the absence of Balmer line despite the rising continuum. 
At the same time, no specific trend is observed has been observed in the X-ray band for the flux and the photon index, between the {\it eROSITA} (2020) and {\it Swift}/XRT (2022-11) observations. 

The non-simultaneity of the optical and X-ray observations between 2004 and 2022 prevents us from properly tracing the relevant timescales or determining how many phase changes the source has undergone. There should have been a drop in the accretion luminosity between 2004 and 2021, explaining what we see in the optical spectra, but, at the same time, it should have been followed by an equal increase of the luminosity to justify the constant level of the X-ray flux.

For this reason, we refrain from linking the changes we see in the optical regime to those seen in the X-rays. Several transitions between high- and low-flux states may have occurred during this period, but they cannot be constrained given the sparse sampling and the intrinsic short-term variability of the X-ray emission.

\subsection{What happened in 2021?}
The 2021-01 spectrum shows a continuum rising towards the blue wavelengths, as in the 2004-04 spectrum, but with an even steeper spectral slope (spectral index of the broken power-law fitting the continuum: index$_{2004-04}$=~-1.8 while index$_{2021-01}$=~-2). This behaviour contrasts with the presence of both broad and narrow Balmer components in the 2004-04 spectrum and the non-detection of the entire H$\beta$ line in the 2021-01 spectrum.

Several explanations can be proposed to account for the variations observed in these two spectra. The difference in the continuum level between the 2004 and 2021 observations could be related either to changes in the accretion state or to the uncertainties on the 2004-04 flux calibration. Assuming the 2004-04 is correctly flux calibrated, the AGN may have experienced several "switch-on-and-off" phases, and the 2021-01 observation may have been taken during the rise of the continuum, just before it reached the BLR and ionised it (considering the lower limit on the BLR size measured from the 2004-04 spectrum, $\approx$6 light-days). This serendipitous timing could explain the total absence of the H$\beta$ line, as neither the BLR nor the NLR would yet have been reached by the newly emerging ionising continuum. However, due to very mild variability in the X-ray flux, this hypothesis should be treated with caution.

It is also worth noting that the Eddington ratio is consistently high in both the 2004-04 ($\lambda_{\rm Edd}$=0.46) and 2021-01 ($\lambda_{\rm Edd}$=0.19) spectra, but also in the \textit{eROSITA} 2020 observation ($\lambda_{\rm Edd}$=0.11). One unexpected aspect is that objects with high $\lambda_{\rm Edd}$ are usually characterised by strong Fe II multiplets \citep{Marziani18a}, which are present in the 2004-04 spectrum but absent in the 2021-01 one. We can state that this non-detection is not related to the S/N ratio (Tab.~\ref{tab:obs}), which is comparable to the other spectra, nor to observational effects. Recent studies support the idea that the production of Fe II lines is governed by photoionisation from the central source \citep{Gaskell22, Ilic23}, a hypothesis motivated by variability analyses \citep{Shapovalova12, Barth13}. For this reason, the physical interpretation for the absence of the iron emission lines could be the same as that suggested for the non-detection of the H$\beta$ broad components: the fortuitous coincidence of having observed the source just before the BLR became fully ionised. As outlined before, this scenario must be carefully taken.

According to the disc-wind scenario \citep{Elitzur14}, an outflow of clouds embedded in a wind originating from the disc is responsible for the formation of the torus and the BLR. In this framework, the BLR is expected to vanish when the AGN luminosity drops below a critical threshold. For this spectrum, the bolometric luminosity exceeds the critical value required for the formation of the broad emission lines (the \citet{Elitzur09} boundary, $\log L = 28.8 - 2\,\log(\lambda_{\rm Edd})$), even though the broad lines are absent.  However, \citet{Jana25} recently showed that the disc-wind model alone is not sufficient to explain all the CL transitions, and that CL AGN can exhibit bolometric luminosities well above the critical threshold predicted by the disc-wind model while still lacking broad emission lines.

Despite the uncertainties and the uncommon spectral features, the differences between the 2004-04 and 2021-01 spectra seem to favour a CS scenario (switch-on-and-off phases), driven by changes in the accretion flow, given that the CL transitions do not appear to be related to the presence of an obscuring medium \citep{Jana25}.

Regarding the second observation of 2021 (Fig.~\ref{fig:NOT_fit}), the continuum level and its slope (index$_{\rm 1, NOT}$=~-2, Fig.~\ref{fig:comparison}), as well as the accretion rate (\ref{tab:lum}), do not show any significant differences with respect to the first observation. As in 2021‑01, neither the broad nor the narrow H$\beta$ components are detected, while both H$\alpha$ components are present.  We cannot determine whether these components were also present in the 2021-01 spectrum due to its shorter wavelength coverage.  One possible interpretation is that we observed the AGN during the very early stages of its turn‑on phase in 2021‑01, which may have led to the ionisation of the BLR seen in 2021‑12 through the reappearance of the broad H$\alpha$ component.

The complete absence of H$\beta$ in the 2021 spectra remains puzzling. The disappearance of the broad H$\beta$ component is consistent with a CL scenario, but the lack of the narrow component is more difficult to explain. One possibility is that the emitting‑line regions are shielded by a puffed‑up disc (often present in NLS1s), similar to what is observed in weak-line quasars with high Eddington ratios \citep{Luo15, Jin17a, Jin17b}. However, if this mechanism affects all optical/UV NLR lines, it should suppress both Balmer and oxygen lines, whereas the latter are clearly visible in the 2021 spectra. We therefore leave open the additional possibility of an obscuring medium absorbing precisely at that wavelength, although no studies have explored this specific scenario. A combination of this phenomenon with the dramatic and rapid transformations occurring in the innermost regions of accreting SMBHs \citep[a CS event,][]{Ricci20} may account for the main spectral variations observed across the epochs.

\subsection{Second phase: 2023}

Lastly, the observed X-ray variability between the two {\it Swift}/XRT observations (2022-11 and 2023-09), characterised by a decrease in flux by a factor of $\sim3$, motivated the subsequent 2023-09 observation, taken shortly after the latest X-ray epoch. The 2023-09 spectra may provide further insight into a new phase of this AGN. The latest optical spectrum shows the reappearance the H$\beta$ line, with both its components. Although J0413-0050 appears to be transitioning toward a IS~1.8 classification, the 5100 \AA \, luminosity, the bolometric luminosity, and the H$\alpha$ integrated flux in the 2023-09 spectrum are still lower than the ones seen in the previous epochs.

The decrease of the flux of the oxygen emission lines in the last optical spectrum can be explained by differences in the slit PA (PA$_{2021-01}$=$\approx -$80$^\circ$, PA$_{2021-12}$=$\approx$45$^\circ$, PA$_{2023-09}$=$\approx$0$^\circ$, Fig.~\ref{fig:pa}), since the slit apertures were the same and the seeing conditions were similar for all the three spectra. In the case of the 2023-09 observation, the slit orientation may have missed additional star-forming regions or the area where the ionised NLR cones or extended-NLR lie. For this reason, we decided to not normalise the spectra to the [O III]$\lambda$4959,5007 fluxes, which is usually done in AGN variability studies to minimise the observational and instrumental effects when comparing spectra from different epochs \citep[e.g.,][and references therein]{Shapovalova19}. On the other hand, the level of the [O III]$\lambda$5007 should remain constant and is expected to vary only over long timescales \citep{Peterson13}, since NLR emission lines are insensitive to rapid continuum flaring because of the large distance, large spatial extent and low gas densities. However, recent studies reported variations in the narrow Balmer components on shorter timescales \citep{Li22}.

As previously discussed, the FWHM of the H$\beta$ broad component (1900 \kms), the continuum luminosity (Tab.~\ref{tab:lum}) and the prominence of the H$\beta$ narrow component compared to the broad one resemble an IS~1.8 classification. The reappearance of the BLR components was not followed by an increase in the accretion rate, possibly suggesting that an obscuration scenario could be responsible for this phase. Variable optical spectra of several IS~1.8 and IS~1.9 galaxies have shown changes in the accretion disc and BLR components \citep[][and references therein]{Goodrich95}. Variations in the profile of the broad emission lines have been interpreted as partial obscuration of the BLR by outflowing dusty gas clumps \citep[e.g.][]{Gaskell18, Zeltyn22}, although this explanation has been considered unlikely in most of the cases \citep{Ricci23}. Due to the mild changes observed in the X-rays and, as already discussed, this cannot be confirmed without N$_{\rm H}$ measurements.

Finally, the exceptionality of this source lies in its CL transition happening at high Eddington ratios, compared to the typical CL behaviour outlined in the Sect.~\ref{sec:intro}. However, recent results suggest that the threshold for such events can reach as low as 1$\%$ of the Eddington ratio \citep[][, and references therein]{Jana25}.

%%%%%%%%%%%%%%%%%%%%%%%%%%%%%%%%%%%%%%%%%%%%%%%%%%%%%%%%%%%%%%%%%%%%%%%%%%%%%%%%%%%%%%%%%%%%%%%%%%%%%%%%%%%
\section{Summary and conclusions}
\label{sec:conclusion}
Non‑simultaneous optical and X‑ray observations of 2MASX J04130709‑0050165 were obtained at several epochs between 2004 and 2023. The first optical spectrum, collected in 2004 within the 6dF survey, led to its classification as a NLS1 galaxy \citep{Chen18}, showing both broad and narrow Balmer components and a high Eddington ratio. In January 2021, the NTT spectrum displayed only forbidden oxygen lines, whereas in December 2021 the NOT observations revealed both components of H$\alpha$. The H$\beta$ line was completely absent in the 2021 spectra, despite the source accreting at a high Eddington rate. The most recent optical spectrum, obtained with UT1 in 2023, indicated an IS~1.8 classification due to the reappearance of both Balmer components.

The available X-ray spectra, obtained in 2020, 2022, and 2023, do not show evidence for a long-term increase in flux; instead, the X-ray emission appears broadly consistent within a factor of a few, with a decrease by a factor of $\sim3$ between the 2022 and 2023 observations. The photon index $\Gamma$ remains consistent within uncertainties across all epochs, indicating no significant spectral evolution. Due to the limited count statistics, it is not possible to constrain the intrinsic column density, $N_{\rm H}$, and therefore we cannot establish whether the source experienced a CT phase. The optical spectral changes observed across the different epochs seem to favour a CS scenario, in which the source underwent multiple “switch-on” and “switch-off” phases. Although this remains the most plausible interpretation, it does not fully account for the complete disappearance of the H$\beta$ line in 2021. Several scenarios may be invoked, particularly to explain the challenging state observed in 2021-01, but a definitive picture will require simultaneous, multi-epoch observations.

To this aim and to obtain a precise determination of the variability timescale of the X-ray flux, as well as to establish a more accurate connection between the flux changes observed in the X-ray and optical ranges, we requested and were granted a 1-year XMM-VLT monitoring programme (P.I. Vietri, A., nr. 94131), consisting of three joint observations between 2024 and 2025. In the X-rays, thanks to the high sensitivity of the EPIC cameras, \xmm\ provided high-quality spectra of the target, in stark contrast to the low net counts we had for the previous X-ray observations. The analysis of the optical and X-rays simultaneous observations will be presented in a forthcoming paper.

In conclusion, this comprehensive analysis enables us to explore in detail the behaviour of a source accreting at a very high rate, as is typical for NLS1s. The Eddington regime appears to be one of the main drivers of the peculiar variability observed in J0413‑0050 over the past two decades. This cadence‑based study allows us to probe the different phases of one of the defining properties of NLS1s. Understanding the physical mechanisms at work during the earliest stages of AGN activity, when the accretion rate can reach its maximum, is a challenging but promising avenue that can be addressed through this approach.

\begin{acknowledgements}
A.V. and M.B. acknowledge the support from the ESO Early-Career Scientific Visitor Programme.
I.V. wants to thank the Swedish Cultural Foundation in Finland for their support. D.I. acknowledges funding provided by the University of Belgrade—Faculty of Mathematics (contract 451-03-66/2024-03/200104) through grants of the Ministry of Education, Science, and Technological Development of the Republic of Serbia. Based on observations collected at the European Southern Observatory under ESO programmes 0104.B-0587(A), 106.21HS and 113.26X0. CR acknowledges support from SNSF Consolidator grant F01$-$13252, Fondecyt Regular grant 1230345, ANID BASAL project FB210003 and the China-Chile joint research fund. This research has made use of the NASA/IPAC Extragalactic Database (NED), which is operated by the Jet Propulsion Laboratory, California Institute of Technology, under contract with the National Aeronautics and Space Administration. This research has made use of the SIMBAD database, operated at CDS, Strasbourg, France. Based on observations made with the Nordic Optical Telescope, owned in collaboration by the University of Turku and Aarhus University, and operated jointly by Aarhus University, the University of Turku and the University of Oslo, representing Denmark, Finland and Norway, the University of Iceland and Stockholm University at the Observatorio del Roque de los Muchachos, La Palma, Spain, of the Instituto de Astrofisica de Canarias. The 2021-12 data were obtained under program ID P64-407. This research has made use of data obtained from the Chandra Data Archive provided by the Chandra X-ray Center (CXC). We thank Dr. Luigi Foschini and Prof. Benjamin Trakhtenbrot for the valuable suggestions. We thank Kostas Valeckas for the support with NOT technical information. We thank Dr. Alessandro Bianchetti for valuable feedback and assistance in improving the clarity and presentation of the manuscript.
\end{acknowledgements}

\bibliographystyle{aa} % style aa.bst
\bibliography{./main.bib} % your references Yourfile.bib

\begin{appendix}
\onecolumn
\label{sec:app}

\section{Host modelling}
\subsection{2004-04 host galaxy modelling}
\label{subsec:6df_app}

\begin{figure*}[h!]
    \centering
    \includegraphics[width=18cm]{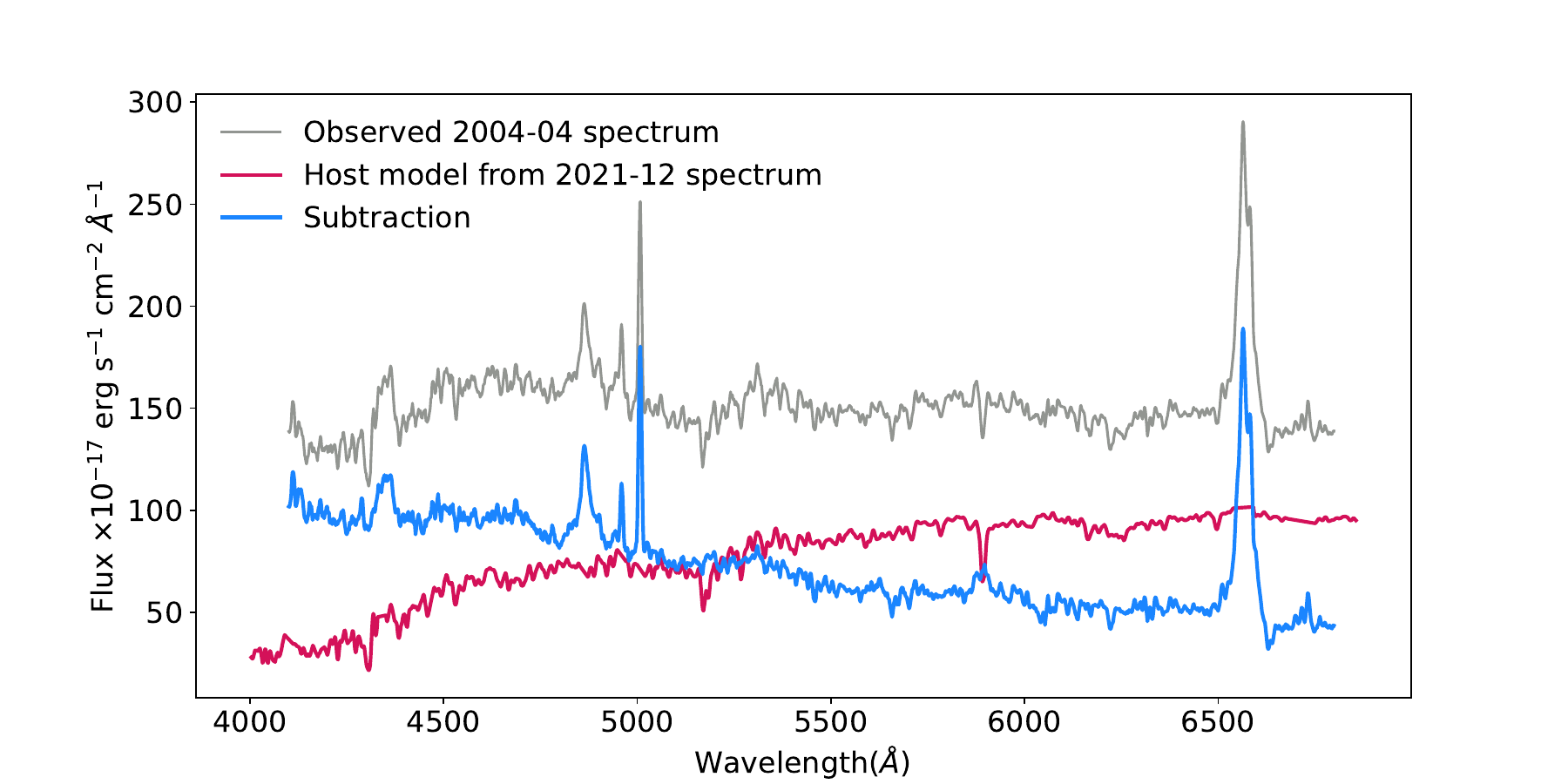}
    \caption{Host galaxy spectrum (2021-12 model, pink magenta), the observed 2004-04 spectrum (grey) and the pure AGN spectrum obtained (bright blue).}
    \label{fig:6df_host}       
\end{figure*}

Since the reconstruction of the host galaxy contribution using eigenspectra did not provide reliable results for the 2004-04 spectrum, we used the host model extracted from the 2021-12 spectrum (see Sect.~\ref{subsec:NOT}). It is reasonable to assume that the host contribution remains constant, as it is not expected to vary significantly over 15-year timescale. To subtract the host model from the 2004-04 spectrum, we rebinned the latter to match the wavelength range of the 2021‑12 spectrum. Fig.~\ref{fig:6df_host} shows the observed spectrum, the host model derived from the 2021-12 data, and the resulting subtraction. The main absorption lines seen in the stellar continuum (G-band at 4304$\AA$, Mg at 5175 $\AA$, Na at 5894$\AA$) disappear from the subtracted spectrum, confirming that it represents a pure AGN spectrum. For this reason, no flux‑scaling correction between the 2004‑04 spectrum and the 2021‑12 host model was required. Once the host contribution is removed, the pure AGN spectrum clearly shows a continuum rising toward the blue wavelengths, as commonly observed in NLS1s \citep{Costantin22}. 

\subsection{2021-01 host galaxy modelling}
\label{subsec:NTT_app}
We reconstructed the host galaxy contribution for the 2021-01 spectrum using all \texttt{fantasy} available eigenspectra, masking the narrow emission lines. The resulting host galaxy model shows stellar absorption features- such as G-band at 4304$\AA$, Mg at 5175 $\AA$ and Na at 5894$\AA$- which, combined with the absence of strong hydrogen absorption lines, resembles a galaxy hosting an older stellar population. This is likely due to the fact that the spectra were extracted from the central region of the galaxy, where the AGN contribution is strongest and where old stars typically dominate.

\begin{figure*}[h!]
    \centering
    \includegraphics[width=18cm]{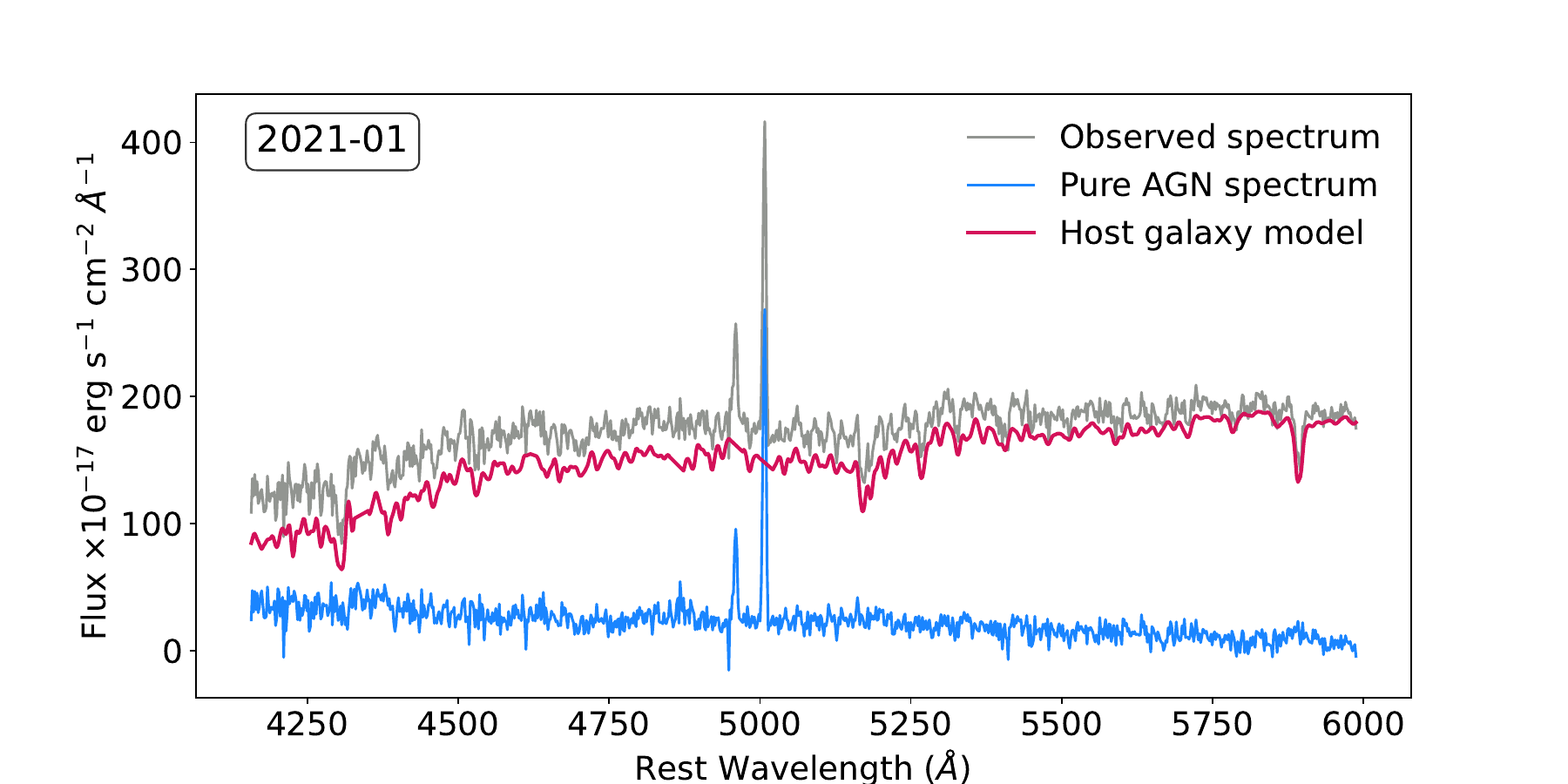}
    \caption{Host galaxy model (magenta pink), observed spectrum (grey) and actual subtraction (bright blue) from the 2021-01 spectrum.}
    \label{fig:NTT_host}       
\end{figure*}

\subsection{2021-12 host galaxy modelling}
\label{subsec:NOT_app}
We performed the host‑galaxy reconstruction for the 2021‑12 spectrum following the same procedure adopted for the 2021‑01 spectrum. Here we show the actual fit of the host model provided by \texttt{fantasy}, which yields the best reduced $\chi^2$. This is the reason why we chose to use this host model to account for the galaxy contribution in the 2004‑04 spectrum.

\begin{figure*}[h!]
    \centering
    \includegraphics[width=18cm]{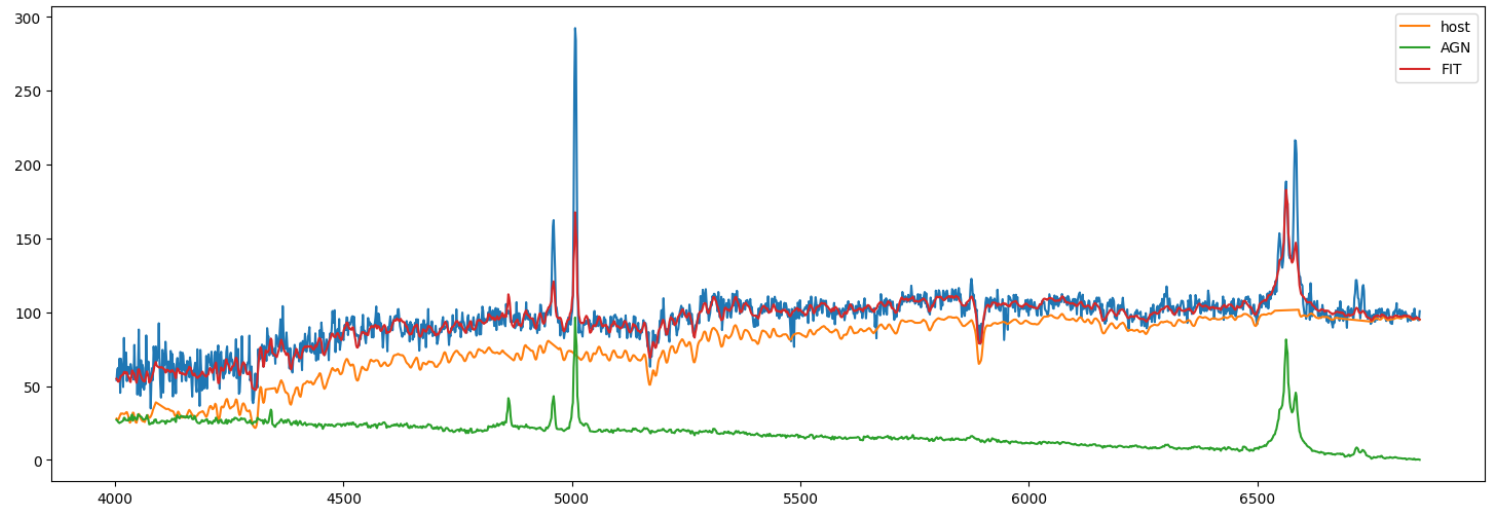}
    \caption{\texttt{fantasy} fit of the host galaxy model for the 2021-12 spectrum. The host model is shown in orange, the extracted AGN spectrum in green, the observed spectrum in blue, and the fit in red.}
    \label{fig:NOT_hostf}       
\end{figure*}

\subsection{2023-09 host modelling}
\label{subsec:UT1_app}
We performed the host‑galaxy reconstruction for the 2023‑09 spectrum following the same procedure adopted for the 2021‑01 spectrum, using \texttt{fantasy}.

\begin{figure*}[h!]
    \centering
    \includegraphics[width=18cm]{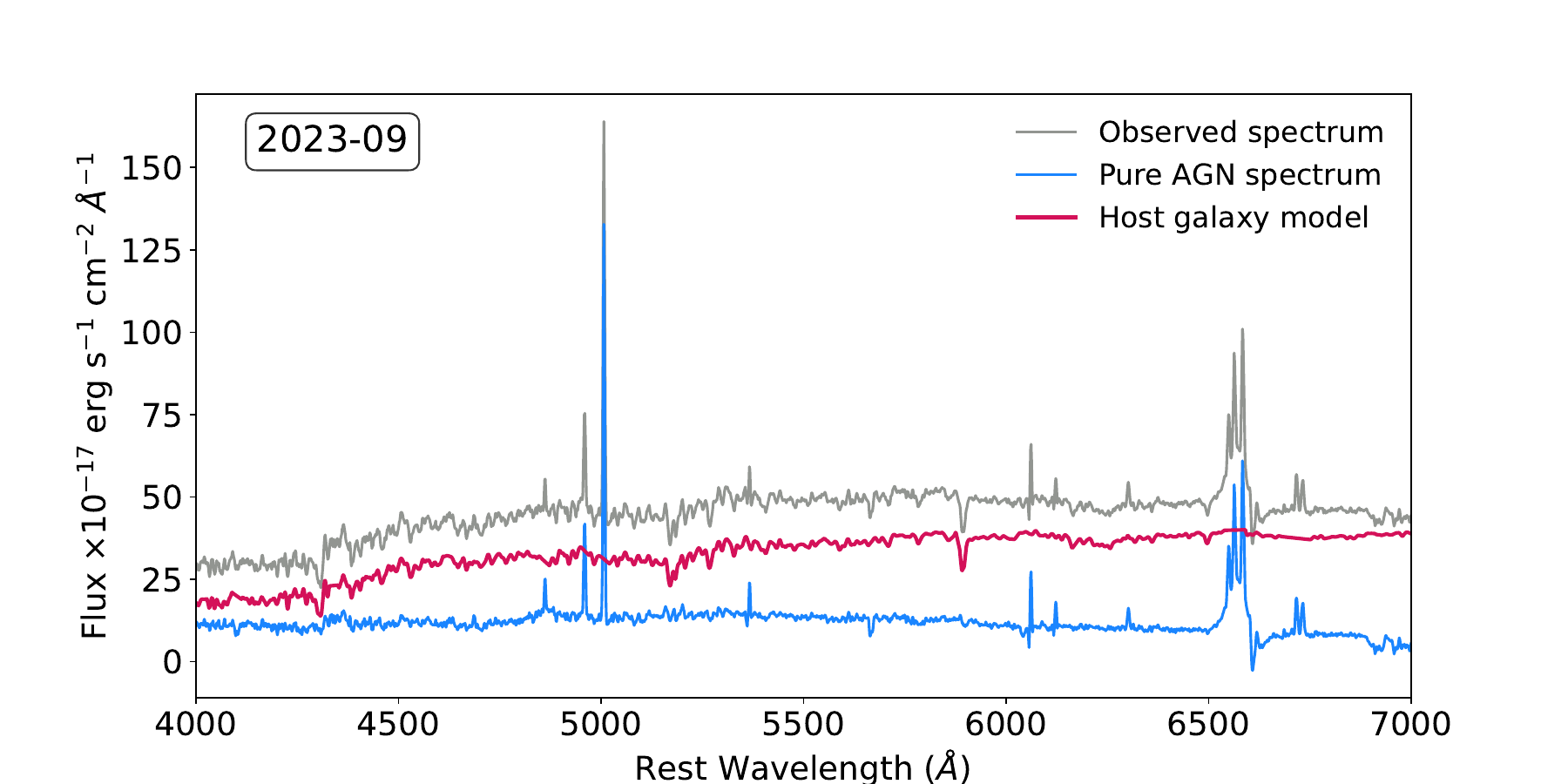}
    \caption{Host galaxy model (magenta pink), observed spectrum (grey) and actual subtraction (bright blue) from the 2023-09 spectrum.}
    \label{fig:UT1_host}       
\end{figure*}

\section{Slit position angles}
\label{sec:pa_app}
\begin{figure*}[h!]
    \centering
    \includegraphics[width=10cm]{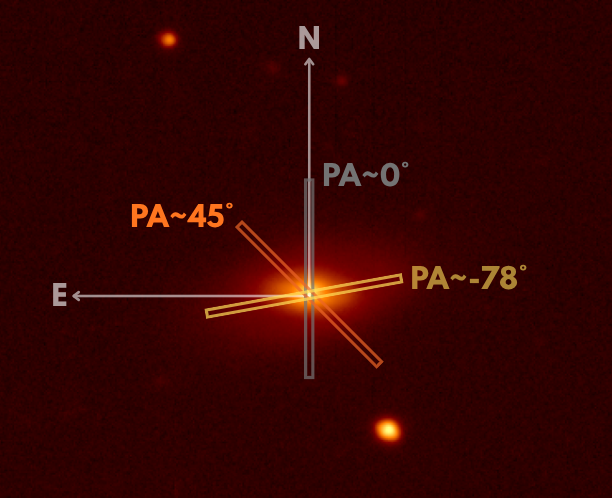}
    \caption{NTT $g-$band image of J0413-0050, oriented North–East (N-E). The PAs of the slits for the different observations are shown. The PAs of the 2021‑01, 2021‑12, and 2023‑09 spectra are indicated in orange, yellow, and grey, respectively.}
    \label{fig:pa}       
\end{figure*}

The choice not to apply absolute [O~III] calibration to the 2021‑01 and 2023‑09 spectra is motivated earlier in the text. Figure~\ref{fig:pa} shows the different slit orientations for each observation.

\section{Light curve}
\label{sec:lc_app}

\begin{figure*}[h!]
    \centering
    \includegraphics[width=18cm]{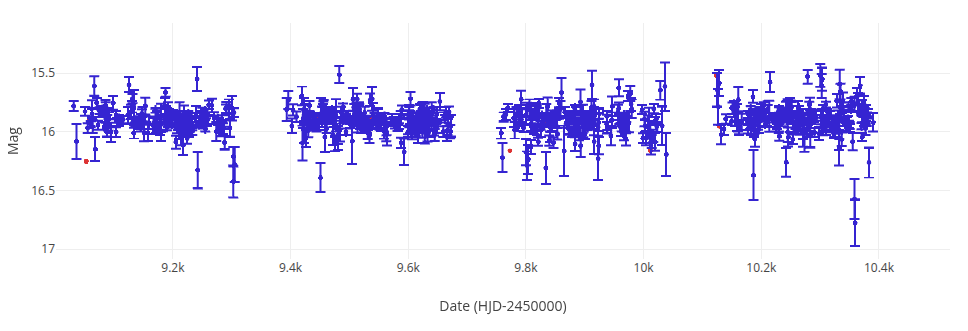}
    \caption{ASAS-SN curve-light 2021-2025}
    \label{fig:lc}       
\end{figure*}

We retrieved the All‑Sky Automated Survey for Supernovae (ASAS‑SN; \url{http://asas-sn.ifa.hawaii.edu/skypatrol/}) light curve, covering the period from 2013 to 2025. It does not show any significant magnitude variations; only mild changes associated with the ‘classical’ AGN variability are present, with no evidence of CL‑related events. It should be noted that the host galaxy may dominate the light curve, potentially preventing dramatic AGN flux changes from being detected. Figure~\ref{fig:lc} shows a zoom‑in of the light curve over the 2021–2025 interval.

\section{Host galaxy}
\label{sec:hg_app}
\label{sec:host}

\begin{figure*}[h!]
    \centering
    \includegraphics[width=18cm]{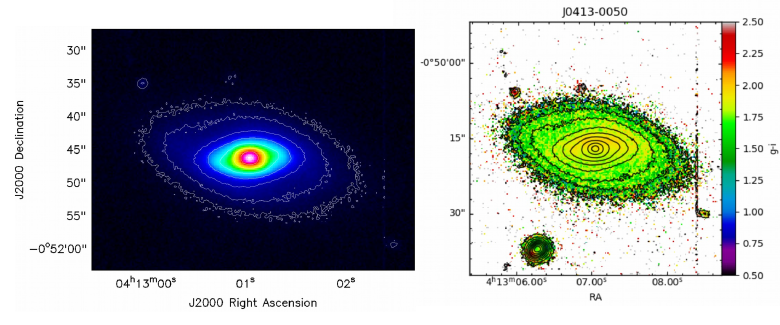}
    \caption{$g-i$ colour map of J0413-0050}
    \label{fig:colourmap}  
\end{figure*}

The optical images of this source were obtained with NTT (proposal ID: 0104.B-0587(A), PI M. Berton) in October 2019. The $g$- and $i$-band observations were carried out using the ESO EFOSC2 (seeing $\approx$1.3"). The exposure time was 300s for both images. We performed a standard reduction using IRAF, including bias and flat‑field correction, followed by alignment, sky subtraction, fringing removal, and combination of the images in each filter.

The $g-i$ colour map is shown in Fig.~\ref{fig:colourmap}. The colour is fairly uniform across the entire galaxy, except for the nucleus. The central region of the map appears yellowish, with $g-i$ $\approx$ 2, a value typically observed in red quasars at this redshift \citep{Klindt19}. The host galaxy image and colour map (Fig.~\ref{fig:colourmap}) reveal a disc structure, suggesting a late-type morphology for J0413-0050, likely hosting an old stellar population in its centre, as also indicated by the absorption lines seen in the host galaxy spectrum.

\end{appendix}

\end{document}